\documentclass[11pt]{article}
\usepackage{jheppub}

\usepackage{amsmath,amssymb,graphicx}
\usepackage{hyperref}
\usepackage{slashed}
\usepackage[font=small,labelfont=bf]{caption}

\usepackage{ifpdf}
\newcommand{\alt}[2]{\texorpdfstring{#1}{#2}}

\newcommand{\be}{\begin{equation}}
\newcommand{\ee}{\end{equation}}
\def\simlt{\stackrel{<}{{}_\sim}}
\def\simgt{\stackrel{>}{{}_\sim}}

\newcommand{\T}{\top}
\newcommand{\dd}{d}
\DeclareMathOperator{\Tr}{Tr}

\DeclareMathOperator{\diag}{diag}
\DeclareMathOperator{\Span}{span}

\usepackage{pifont}
\newcommand{\cmark}{\ding{51}}%
\newcommand{\xmark}{\ding{55}}%
\newcommand{\cc}[1]{\mathcal{#1}}

\newcommand{\bb}[1]{\mathbb{#1}}

\usepackage[usenames,dvipsnames]{xcolor}

\title{\centering Evidence for a Lattice Weak Gravity Conjecture}

\author[a,b]{Ben Heidenreich,}
\author[a]{Matthew Reece,}
\author[a]{and Tom Rudelius}

\affiliation[a]{Department of Physics, Harvard University, Cambridge, MA
  02138, USA}
\affiliation[b]{Perimeter Institute for Theoretical Physics, Waterloo, Ontario, Canada N2L 2Y5}  
  
\emailAdd{b.j.heidenreich@gmail.com}
\emailAdd{mreece@physics.harvard.edu}
\emailAdd{rudelius@physics.harvard.edu}

\abstract{The Weak Gravity Conjecture postulates the existence of superextremal charged particles, i.e.~those with mass smaller than or equal to their charge in Planck units. We present further evidence for our recent observation that in known examples a much stronger statement is true: an infinite tower of superextremal particles of different charges exists. We show that effective Kaluza-Klein field theories and perturbative string vacua respect the Sublattice Weak Gravity Conjecture, namely that a finite index {\em sublattice} of the full charge lattice exists with a superextremal particle at each site. In perturbative string theory we show that this follows from modular invariance. However, we present counterexamples to the stronger possibility that a superextremal particle exists at {\em every} lattice site, including an example in which the lightest charged particle is subextremal. The Sublattice Weak Gravity Conjecture has many implications both for abstract theories of quantum gravity and for real-world physics. For instance, it implies that if a gauge group with very small coupling $e$ exists, then the fundamental gravitational cutoff energy of the theory is no higher than $\sim e^{1/3} M_{\rm Pl}$.}

\begin{document}
\maketitle

\section{Introduction}\label{sec:intro}

Although the landscape of low-energy effective field theories that can be consistently coupled to gravity is vast, it is likely only a small fraction of the set of all possible effective theories.  Theories that cannot be consistently completed into gravitational theories are said to reside in the ``Swampland" \cite{Vafa:2005ui,Ooguri:2006in}.  Precisely delineating the boundaries of the Swampland is necessary for deriving precise predictions from string theory and bringing quantum gravity into contact with experiment.

To date, one of the sharpest and most useful criteria for distinguishing theories in the Swampland from those in the landscape is the Weak Gravity Conjecture \cite{ArkaniHamed:2006dz}:\\

\noindent
{\bf The Weak Gravity Conjecture (WGC):}  In any $U(1)$ gauge theory coupled to gravity, there must exist a superextremal particle.\\

\noindent
Here and throughout this paper, we use the term ``superextremal" to describe a particle $p$ whose charge-to-mass ratio $|\vec{q}_p| /m_p$ is greater than \emph{or equal to} that of a large, non-rotating, extremal black hole of proportional charge:
\begin{equation}
 \frac{|\vec{q}_p|}{m_p} \geq \frac{|\vec{Q}|}{M}  \Big|_{\rm ext} \,,~~\vec{q_p} \propto \vec{Q} ~~~ \Leftrightarrow~~~ \textrm{$p$ is superextremal.}
\label{wgcbound1}
\end{equation}
A particle whose charge-to-mass ratio does not obey this bound is said to be subextremal.  The WGC also has a magnetic dual, which states that the magnetic charge-to-mass ratio of a magnetic monopole should be greater than or equal to that of an extremal, magnetically charged black hole.  Identifying the classical monopole radius as a cutoff $\Lambda$ on the validity of local effective field theory, this implies a bound $\Lambda \lesssim e M_{\rm Pl}$.\footnote{To be precise, this is the scale at which additional charged particles must appear, such as KK modes or string modes. Incorporating these modes appropriately, the effective field theory description can be extended as high as the scale $e^{1/3} M_{\rm Pl}$, as discussed in~\S\ref{subsec:UVcutoffs}.}

Why should the WGC be true?  The conjecture was originally motivated by the desire to avoid stable black holes: in a $U(1)$ gauge theory that obeys the WGC, charged black holes can typically decay (perhaps at threshold) by emitting a superextremal particle.\footnote{The WGC is often defined to be the statement that black holes should be able to decay.  However, stable black holes can exist even in a theory with a superextremal particle~\cite{Bachlechner}.  We will see an example of a tree-level spectrum that satisfies the WGC yet contains an infinite tower of stable black holes in section \ref{ssec:examples}.}  In a theory that violates the WGC, there will be an infinite tower of stable, extremal black holes. This tower of states does not violate any known entropy bounds, and it is unclear that stable black hole states present any fundamental problem.  Several works have sought to motivate the WGC from different perspectives \cite{ArkaniHamed:2006dz, banks:2006mm, cheung:2014ega, delaFuente:2014aca, nakayama:2015hga, harlow:2015lma, Horowitz:2016ezu}, but none have successfully established the bound (\ref{wgcbound1}) in general.  Thus, a convincing bottom-up argument for the WGC remains elusive.  Nonetheless, the WGC has been shown to hold in a large class of examples in string theory \cite{ArkaniHamed:2006dz}, and to date there are no known counterexamples.  Examples and internal consistency checks arising from string theory and Kaluza-Klein theory have proven especially useful for clarifying and ruling out possible variants of the WGC \cite{Heidenreich:2015nta}, and they will play a vital role in this paper as well.

From a phenomenological perspective, the WGC is especially interesting due to its ability to constrain models of large-field axion inflation \cite{Rudelius:2014wla,Bachlechner:2014gfa,delaFuente:2014aca,Rudelius:2015xta,Montero:2015ofa,Brown:2015iha,
Bachlechner:2015qja,Hebecker:2015rya,Brown:2015lia,Junghans:2015hba,Heidenreich:2015wga,
Palti:2015xra,Kooner:2015rza,kappl:2015esy,choi:2015aem, Ibanez:2015fcv,Hebecker:2015zss, conlon:2016aea,baume:2016psm,Heidenreich:2016jrl}.
These no-go results, however, have loopholes that permit models of large-field axion inflation consistent with the WGC.  The loopholes could be closed if additional, stronger variants of the conjecture hold, making the identification of any correct ``strong forms" of the WGC a crucial task from the perspective of inflationary model building.  Two candidate strong forms of the WGC were written down in the original work of \cite{ArkaniHamed:2006dz}.  The first states that the particle \emph{of smallest charge} should be superextremal.  The second strong form holds that the \emph{lightest} charged particle in the spectrum should be superextremal. Below, we will see that these two candidate strong forms are both false.

Our discussion thus far has focused on theories with a single $U(1)$ gauge field.  In practice, however, one expects many $U(1)$s in a low-energy effective theory arising from a UV-complete theory of quantum gravity.  The mildest form of the WGC generalizes straightforwardly: in any rational direction in charge space, there should exist some (possibly multiparticle) superextremal state.  Here, the ``mass" of a multiparticle state is simply the sum of the masses of the individual particles comprising it and is equivalent to the ADM energy carried by the state provided spacetime is asymptotically flat, so any interactions between the particles die off at infinity.  A ``rational direction" in charge space is any ray in charge space that contains a point on the lattice.  This statement has a nice geometric interpretation known as the ``Convex Hull Condition" \cite{cheung:2014vva}. Moreover, it generalizes easily to non-Abelian gauge groups in terms of their Cartan subgroup;\footnote{Although the WGC is trivially satisfied by the gauge bosons for an unbroken non-Abelian gauge group, this is no longer true when we discuss lattice strong forms of the WGC below.} the connection to the Abelian case is evident upon reducing the non-Abelian theory on a circle with generic Wilson lines along it.

The strong forms of the conjecture, however, are more difficult to generalize to theories with multiple $U(1)$s~\cite{Heidenreich:2015nta}.
One conjectured strong form for theories with multiple $U(1)$s is that the lightest (possibly multiparticle) state in any rational direction in charge space should be superextremal \cite{Brown:2015iha}.  An even stronger form is the ``Lattice Weak Gravity Conjecture" \cite{Heidenreich:2015nta}:
\\

\noindent
{\bf The Lattice Weak Gravity Conjecture (LWGC):}  In any gauge theory coupled to gravity, any spot on the charge lattice $\vec{q}$ consistent with Dirac quantization should contain a (possibly unstable) superextremal particle.\\

\noindent
This strong form implies the previous one, but it has the advantage of behaving nicely under Kaluza-Klein reduction on a circle \cite{Heidenreich:2015nta}.  In this paper, however, we will see that even the LWGC is incompatible with Kaluza-Klein reduction on more complicated manifolds: Kaluza-Klein modes do not always saturate the black hole extremality bound in their direction in charge space, leaving some spots on the charge lattice devoid of a superextremal particle.  In the examples we have studied, however, the extremality bound is still saturated by all of the KK modes in a \emph{sublattice} of the full charge lattice; for instance, we show that this is true for KK reductions of pure gravity with no cosmological constant.  This motivates the ``Sublattice Weak Gravity Conjecture" (sLWGC), which holds that a superextremal particle should exist at every spot in a finite index sublattice of the full charge lattice.  Strong evidence for the sLWGC comes from perturbative string theory, where the conjecture is related to modular invariance. The conjecture can also be verified explicitly in a large class of theories produced via compactification of the type II and heterotic string on toroidal orbifolds.  In some of these theories, the lightest charged state in the spectrum is actually subextremal and the theory violates each of the aforementioned strong forms, though it satisfies the sLWGC. The case of Type IIB string theory on AdS$_5 \times$ S$^5$ gives another example in which the sLWGC is satisfied.

The sLWGC has a number of interesting consequences for effective field theories.  Like the magnetic form of the WGC, the sLWGC implies a cutoff on effective field theory due to the presence of an infinite tower of massive particles.  Applied to stable particles, the sLWGC implies that a finite index sublattice of the charge lattice must contain superextremal multiparticle states.\footnote{However, this statement is only a little stronger than the Convex Hull Condition. In particular, if the Convex Hull Condition is satisfied by a finite number of charged particles, then this ``stable particle sLWGC'' follows.} The sLWGC also constrains models of axion inflation, though we postpone a detailed discussion to a later work \cite{Liam}.

Before proceeding to a detailed analysis of examples, let us clarify our approach in this paper. The WGC can be studied from several perspectives. A bottom-up argument, showing that some form of the WGC is necessary for the consistency of any gravitational theory, would be ideal. In the absence of such arguments, we believe that progress can be driven from other considerations. Internal consistency checks provide one important tool for honing our conjecture; this was our primary approach in \cite{Heidenreich:2015nta}, where we argued that consistency of the WGC under compactifications calls for the existence of {\em towers} of charged states. Top-down ``data'' from string theory can also provide guidance, by showing us how concrete, consistent quantum gravity theories behave. That is the approach that we will mostly take in the current paper. Although such an approach can never lead to a proof of a general statement about quantum gravity, it can falsify such statements---indeed, we will falsify several different strong forms  of the WGC as we proceed. It can also offer circumstantial evidence in favor of a conjecture. Although the original LWGC is falsified by our investigations, the sLWGC is not, and remains consistent with our previous arguments from internal consistency that the WGC should call for a tower of particles. We believe that all three approaches---bottom-up, top-down, and internal consistency arguments---play an important role in making progress.\\

{\bf Note added}: we have learned that the modular invariance argument we present in \S\ref{subsec:modularinvariance} has been independently derived by Montero, Shiu, and Soler and applied to the Weak Gravity Conjecture in AdS$_3$ \cite{Montero:2016tif}. This application provides an intriguing additional piece of evidence for the broad validity of a lattice form of the WGC.

\section{Lattice Weak Gravity and Kaluza-Klein Reduction}\label{sec:KK}

Kaluza-Klein theory provides a natural realization of a gauge theory coupled to gravity. The Lattice Weak Gravity Conjecture (LWGC) of~\cite{Heidenreich:2015nta} was motivated in part by the behavior of toroidal Kaluza-Klein compactifications, which we now review to establish notation.

\subsection{Toroidal compactification}

Consider a $D$-dimensional pure gravity theory compactified on an $k$-torus $T^k$, parameterized by the angles $\theta_i \cong \theta_i + 2 \pi$, $i=1,\ldots, k$. We take the general metric ansatz
\be
d s^2_D = |\varphi|^{-\frac{1}{d-2}} g_{\mu \nu} dx^\mu dx^\nu+R^2 \varphi_{i j} (d \theta^i + A^i_{\mu} d x^\mu) (d \theta^j + A^j_{\nu} d x^\nu) \,,
\ee
where $g_{\mu\nu}$ denotes the $d=D-k$ dimensional metric, $\varphi_{ij}$ the $k(k+1)/2$ metric moduli with $|\varphi|\equiv \det \varphi_{i j}$, $A^i_{\mu}$ the $k$ graviphotons, and $R^2$ the overall scale of the compactification. We expand the fields $g_{\mu \nu}$, $\varphi_{i j}$, and $A^i_\mu$ in plane waves on $T^k$
\be
\varphi_{i j} (x,\theta) = \sum_{\vec{n} \in \bb{Z}^k} \varphi_{i j}^{(n_1,\ldots,n_n)}(x)\, e^{i \sum_i \theta^i n_i} \,,
\ee
and likewise for $g_{\mu \nu}$ and $A^i_{\mu}$. The dimensionally reduced action for the zero modes $\widehat{\varphi} \equiv \varphi^{(\vec{0})}$ takes the form:
\be \label{eqn:KKeffectiveaction}
S = \frac{1}{2 \kappa_d^2} \int d^d x \sqrt{-\widehat{g}} \left( \cc{R}[\widehat{g}] - \frac{1}{4} \left[\widehat{\varphi}^{i k} \widehat{\varphi}^{j l} + \frac{1}{d-2} \widehat{\varphi}^{i j} \widehat{\varphi}^{k l} \right] \nabla \widehat\varphi_{i j}\cdot\nabla \widehat\varphi_{k l} - \frac{R^2}{2} |\widehat\varphi|^{\frac{1}{d-2}} \widehat\varphi_{i j} \widehat F^i\cdot \widehat F^j \right) \,,
\ee
where $\widehat{\varphi}^{i j} \equiv (\widehat{\varphi}^{-1})^{i j}$ and $\omega_p \cdot \chi_p \equiv \frac{1}{p!}\, \omega_{\mu_1 \ldots \mu_p} \chi^{\mu_1 \ldots \mu_p}$. The $U(1)^k$ gauge symmetry corresponds to translations on $T^k$, hence the zero modes are neutral whereas the remaining KK modes carry mass and charge determined by their mode numbers:\footnote{To simplify this formula, we assume without loss of generality that $\det \widehat{\varphi}_{i j} = 1$ in the background of interest, absorbing the overall volume of $T^k$ into the definition of $R$. In general, the correct formula is $m^2 = |\widehat{\varphi}|^{-\frac{1}{d-2}} \widehat{\varphi}^{i j} n_i n_j /R^2$.}
\begin{align} \label{eqn:KKmasses}
m^2 &= \widehat{\varphi}^{i j} n_i n_j /R^2 \,, & Q_i &= n_i\,.
\end{align}

Decomposing $\widehat{\varphi}_{i j} = \delta_{a b} e^a_i e^b_j$ using the vielbein $e^a_i$, we find that the metric moduli space is $\frac{GL^+(k,\bb{R})}{SO(k)\times SL(k,\bb{Z})}$ where $GL^+(k,\bb{R})$ denotes the connected component $\det e^a_i >0$ and the two denominator factors act on the left and right, respectively. Note that $GL^+(k,\bb{R})$ is promoted to a symmetry in the classical effective action for the zero modes, despite acting non-trivially on the massive charged spectrum. This simplifies the problem of constructing black hole solutions, as we can set $\widehat{\varphi}_{i j} \to \delta_{i j}$ asymptotically and take $Q_2 = \ldots = Q_k = 0$ without loss of generality. The scalar equation of motion simplifies to
\be
\nabla\cdot( \widehat{\varphi}^{i k} \nabla \widehat{\varphi}_{j k}) = R^2\, |\widehat{\varphi}|^{\frac{1}{d-2}} \widehat{\varphi}_{j k} F^i \cdot F^k \,,
\ee
hence we can consistently truncate $\widehat{\varphi}_{i j} = \diag(e^{-\lambda}, 1, \ldots, 1)$ and $F_j = (F,0,\ldots,0)$. The problem now reduces to that of finding black hole solutions in pure gravity reduced on a circle. Following e.g.~\cite{Heidenreich:2015nta} and generalizing the result using $GL^+(k,\bb{R})$ invariance, we obtain the extremality bound:
\be \label{eqn:KKextremality}
M^2 \ge \widehat{\varphi}^{i j} \frac{Q_i Q_j}{R^2} \,.
\ee
We conclude that the KK modes~(\ref{eqn:KKmasses}) are exactly extremal. Since there is (at least) one KK mode for every possible charge, the LWGC is satisfied.

Despite this simple structure, which remarkably persists in a number of simple string theory examples such as the ten dimensional heterotic string, its toroidal compactifications, and certain simple orbifolds thereof, we now show that the LWGC does not hold in general KK compactifications (nor in all string compactifications). To do so, we generalize to non-toroidal compactifications. In a pure gravity theory, we can compactify on any smooth Ricci-flat manifold.\footnote{We assume a vanishing cosmological constant.} Even if the compact manifold is completely flat, it is in general a toroidal orbifold rather than simply a torus. The resulting Kaluza-Klein theory is somewhat richer than the toroidal case. To illustrate this we now consider a simple $T^3/(\bb{Z}_2\times \bb{Z}_2')$ orbifold example, later generalizing to arbitrary Ricci flat manifolds.

\subsection{A simple orbifold} 
\label{subsec:simpleexample}

Consider a compactification on the smooth manifold obtained by quotienting $T^3$, parameterized by three angles $\theta_{w,y,z}$ as above, by the freely acting $\bb{Z}_2 \times \bb{Z}'_2$ group\footnote{Note that $T^3 / \bb{Z}_2 = (T^{3})'$ is another torus, hence this orbifold can be written more concisely as $(T^3)'/\bb{Z}_2'$. However, the representation $T^3 / (\bb{Z}_2 \times \bb{Z}_2')$ is more convenient, as it makes manifest the massless and massive KK photons.}
\begin{align}
\begin{split}
\bb{Z}_2: \quad & \theta_w \mapsto \theta_w + \pi,\, \theta_y \mapsto \theta_y + \pi, \\
\bb{Z}'_2: \quad & \theta_w \mapsto -\theta_w,\, \theta_z \mapsto \theta_z + \pi.
\end{split}
\end{align}
Note that the $\bb{Z}_2'$ acts as a ``roto-translation'': a rotation combined with a translation in a different direction. Unlike pure rotations, roto-translations can be freely acting, as required for a smooth orbifold geometry.

The original toroidal compactification contains 3 gauge fields associated with Kaluza-Klein momentum in the three directions; we will denote them $W_\mu$, $Y_\mu$, and $Z_\mu$. The action of $\bb{Z}'_2$ projects the first of these fields out of the spectrum, so the orbifold compactification has a $U(1)_Y \times U(1)_Z$ Kaluza-Klein gauge symmetry. Some Kaluza-Klein modes of $W_\mu$ remain in the spectrum and play an important role, as we will see below.

The mode decomposition of a field $\phi$ on the $T^3$ is given by
\be
\phi(x^\mu, \theta_w, \theta_y, \theta_z) = \sum \phi_{n_w, n_y, n_z}\!(x^\mu)\, e^{i n_w \theta_w +i  n_y \theta_y + i n_z \theta_z}.
\ee
The group action imposes the identifications
\begin{align}
\begin{split}
\bb{Z}_2: \quad \phi_{n_w,n_y,n_z}(x) &= \left(-1\right)^{n_w + n_y} \phi_{n_w,n_y,n_z}\!(x), \\
\bb{Z}'_2: \quad \phi_{n_w,n_y,n_z}(x) &= \left(-1\right)^{n_z} \sigma(\phi) \phi_{-n_w,n_y,n_z}\!(x).
\end{split}
\end{align}
Here $\sigma(\phi)$ denotes an additional sign that may arise depending on the nature of the field $\phi$. In particular, for the graviton, the metric moduli $\varphi_{w w}, \varphi_{y y}, \varphi_{y z}$ and $\varphi_{z z}$, and the gauge fields $Y_\mu$ and $Z_\mu$, this factor is simply $1$. However, we have $\sigma(W_\mu) = -1$, because $W_\mu$ arises from the $g_{\mu w}$ modes of the metric which acquire an extra sign under the action of $\bb{Z}'_2$. The metric moduli $\varphi_{w y}$ and $\varphi_{w z}$ have $\sigma = -1$ for the same reason.

Now we ask: for which sites of the charge lattice do we find a superextremal mode? The charge lattice is defined by integer values of $(n_y, n_z)$. Depending on whether the charges are even or odd, we have three cases:
\begin{itemize}
\item {\bf Both charges even.} In this case, modes of the graviton with $n_w = 0$ will saturate the extremality bound. \cmark
\item {\bf Odd $n_z$, even $n_y$.} If $n_z$ is odd and $n_w = 0$, a mode will be projected out by $\bb{Z}'_2$ {\em unless} we have $\sigma(\phi) = -1$. Hence, for these sites in the charge lattice, graviton KK modes are subextremal but KK modes of the (broken) gauge field $W_\mu$ saturate the extremality bound (as do those of the massive moduli $\varphi_{w y}$ and $\varphi_{w z}$). \cmark
\item {\bf Odd $n_y$.} For all fields, the first $\bb{Z}_2$ ensures that unprojected modes with odd values of $n_y$ must also have an odd value of $n_w$. In this case the mode gets an additional contribution $(n_w/R_w)^2$ to its mass squared. No superextremal modes exist at these points of the charge lattice. \xmark
\end{itemize}
We see that the sites in the charge lattice for which superextremal particles exist are those with $n_y$ even and $n_z$ arbitrary. This is a proper sublattice of the full charge lattice, so the Lattice Weak Gravity Conjecture fails, but a related statement is true. This leads naturally to a revised Lattice Weak Gravity
 Conjecture:\footnote{The case $D=4$ is special because the gauge coupling runs logarithmically in the infrared in the presence of light charged particles. When there are massless charged particles, as at a conifold transition in moduli space of type II string theory on an empty Calabi-Yau threefold, the gauge coupling flows to zero in the infrared, yet there is still a (logarithmically renormalized) long range force. In these cases, it may be necessary to revisit what is meant by a ``superextremal'' charged particle. We defer further consideration of this to future work~\cite{Alim2017}; for $D>4$ the logarithmic running is absent and this issue does not arise.}\\

\noindent
{\bf The Sublattice Weak Gravity Conjecture (sLWGC):} For a theory with charge lattice $\Gamma$, there exists a sublattice $\Gamma_{\rm ext} \subseteq \Gamma$ of finite coarseness such that for each $\vec{q} \in \Gamma_{\rm ext}$, there is a (possibly unstable) superextremal particle of charge $\vec{q}$.\\

Here we define the ``coarseness" of the sublattice $\Gamma_{\rm ext} \subseteq \Gamma$\footnote{Here ``ext'' is a shorthand for (super)extremal.} to be the smallest integer $N$ such that $N \vec{q} \in \Gamma_{\rm ext}$ for any $\vec{q} \in \Gamma$.  In particular, this implies that the sublattice of superextremal particles has the same dimension as the full charge lattice.  It also implies that $\Gamma_{\rm ext}$ has a finite index with respect to $\Gamma$, hence the quotient $\Gamma/\Gamma_{\rm ext}$ is a finite group. Note that the {\em index} of the sublattice is the order of $\Gamma/\Gamma_{\rm ext}$, while the \emph{coarseness} is the least common multiple of the element orders within $\Gamma/\Gamma_{\rm ext}$ (equal to the largest element order because $\Gamma/\Gamma_{\rm ext}$ is Abelian). The two are not equal unless $\Gamma/\Gamma_{\rm ext}$ is cyclic, and in general coarseness $\leq$ index $\leq$ coarseness$^k$ for a $k$-dimensional lattice. The finite group $\Gamma/\Gamma_{\rm ext}$ might play an interesting role in black hole physics and in defining the correct form of the Weak Gravity Conjecture, but we leave further speculation on this point for the future.

After arguing that the sLWGC is a generic feature of KK compactification, we will present top-down evidence for the sLWGC, as well as highlighting UV-complete examples which violate all of the strong forms of the WGC discussed in section~\ref{sec:intro} but which satisfy the sLWGC.  

\subsection{Arbitrary Ricci flat manifolds} \label{subsec:Ricciflat}

We now show that the main features of the above example generalize to arbitrary compactifications on Ricci flat manifolds.

We begin by considering a generic smooth toroidal orbifold $T^k / G_0$, from which the generalization to arbitrary Ricci flat manifolds is remarkably simple. Here $G_0$ is a freely acting finite group of translations and rotations on $T^k$. The purely translational subgroup of $G_0$ is normal, and we can take it to be trivial without loss of generality by reducing the size of the $T^k$ fundamental domain. Since the action is free, the non-trivial elements of $G_0$ are all roto-translations.

Consider the projection map
\be
\Pi: V \mapsto \frac{1}{|G_0|} \sum_{g \in G_0} g V 
\ee
on the tangent space. The image and kernel of $\Pi$ (the latter equal to the image of $1-\Pi$) define complementary subspaces of the space of $T^k$ Killing vector $\partial_{\theta_i}$. The former is the dimension $p\le k$ space of $G_0$-invariant Killing vectors whereas the later is the dimension $k-p$ ``null'' space of Killing vectors with no $G_0$ invariant component. Because $G_0$ acts on the tangent space as a subgroup of $GL(k,\bb{Z})$, each of these spaces admits a basis consisting of rational linear combinations of the $T^k$ Killing vectors $\partial_{\theta_i}$.\footnote{To form such a basis for the image of $\Pi$, apply $\Pi$ to the basis $\partial_{\theta_i}$ and eliminate linearly dependent vectors one by one until none remain. To form such a basis for the kernel of $\Pi$, repeat the above, replacing $\Pi \to 1-\Pi$.} Therefore, the orbits of the Killing vectors close and the invariant and null subspaces generate subtori $T^p \subset T^k$ and $T^{k-p} \subset T^k$, respectively. We have
\be
T^k = \frac{T^p \times T^{k-p}}{H} 
\ee
for some discrete translation group $H$. By construction, the elements of $G_0$ act by simultaneous translations on $T^p$ and rotations or roto-translations on $T^{k-p}$. In particular, $T^{k-p}$ tangent vectors cannot acquire a component along $T^p$ under the action of $G_0$, since the latter would not be null.

Quotienting $T^{k-p}$ by the roto-translations which act trivially on $T^p$ and combining the remaining elements of $G_0$ with $H$, we obtain the representation:
\be \label{eqn:orbifoldfactorization}
\frac{T^k}{G_0} = \frac{T^p \times \widehat{M}_{k-p}}{G} \,,
\ee
where $\widehat{M}_{k-p}$ is a smooth toroidal orbifold and $G$ acts by simultaneous translations on $T^p$ combined with discrete isometries of $\widehat{M}_{k-p}$, such that non-trivial elements of $G$ act non-trivially on each factor and there are no $G$-invariant Killing vectors of $\widehat{M}_{k-p}$. As a consequence of the latter, flat $G$-invariant metrics on this space contain no cross terms between $T^p$ and $\widehat{M}_{k-p}$, and the factorization~(\ref{eqn:orbifoldfactorization}) is geometric as well as topological.

We now argue that the sLWGC holds in the $d$-dimensional Kaluza-Klein theory arising from $D=d+k$ dimensional pure gravity compactified on $T^k/G_0$. The gauge group is $U(1)^p$, arising from the $G$ invariant Killing vectors generating $T^p$, where the gauge charge is given by the mode numbers $\vec{Q} = (n_1, \ldots, n_p)$ as usual. The $G$ action on the KK modes is\footnote{The KK modes can be organized into eigenstates of $G$. In such a basis, we do not need to consider mixing between different modes $\nu \ne \nu'$ under the action of $G$.}
\be
g: \phi_{\vec{n},\nu} \mapsto e^{2 \pi i \vec{g}\cdot \vec{n}} \sigma_\nu^{(\phi)}(g) \phi_{\vec{n},\nu} \,,
\ee
where $\nu$ labels the mode expansion on $\widehat{M}_{k-p}$, the map $g \mapsto \vec{g}$ is an injective homomorphism from $G$ to $T^p \cong \bb{R}^p/\bb{Z}^p$ and $\sigma_\nu^{(\phi)} : G \to U(1)$ is a mode and field-dependent phase factor.

The group $G$ defines a lattice $\Gamma_G \supseteq \bb{Z}^p$ via $T^p / G \cong \bb{R}^p / \Gamma_G$, where each element of $g \in G$ corresponds to a coset $\bb{Z}^p+\vec{g}$ within $\Gamma_G$. The dual lattice $\Gamma_G^\ast$ is contained in the charge lattice, $\Gamma \equiv \bb{Z}^p$, and each site in $\Gamma_G^\ast$ is populated with extremal particles, as follows: consider any one of the massless fields $\phi$ which survives the $G$ projection, i.e.\ with $\sigma_{0}^{(\phi)}=1$. (There are always fields in this category, since the graviton is massless.) The $G$-invariant KK modes of $\phi$ include those with $\nu = 0$ and $\vec{n} \in \Gamma_G^\ast$, which have charge $\vec{Q} = \vec{n}$ and mass
\be
m^2 = \hat{\varphi}^{i j} Q_i Q_j/R^2 \,.
\ee
In general, the KK modes satisfy
\be \label{eqn:KKineq}
m^2 \ge \hat{\varphi}^{i j} Q_i Q_j/R^2 \,,
\ee
where the equality is saturated for a given charge if and only if there is a $G$-invariant mode with $\nu=0$. Since $\sigma_{\nu}^{(\phi)}(h) = 1$ for all $h \in H$, such modes are confined to the sublattice $\Gamma_H^\ast \subseteq \Gamma$, which is proper if $H$ is non-trivial.

To show that the sLWGC holds, it remains to be shown that~(\ref{eqn:KKineq}) is the black hole extremality bound. The low energy effective action is~(\ref{eqn:KKeffectiveaction}) with the fields $\varphi_{i \alpha}$, $A^\alpha$, and some of the $\widehat{M}_{k-p}$ metric moduli $\varphi_{\alpha \beta}$ removed,\footnote{The orbifold also affects the normalization of $\kappa_d^2$ v.\ $\kappa_D^2$ due to the reduced volume of the orbifold, but this is an overall factor which does not affect the extremality bound.} where $i=1,\ldots,p$ and $\alpha= p+1,\ldots, k$. For black holes with $Q_\alpha = 0$, these fields were already truncated in the relevant black hole background, and the extremality bound~(\ref{eqn:KKextremality}) is unchanged and matches~(\ref{eqn:KKineq}).

Thus, the sLWGC is satisfied with the extremal sublattice $\Gamma_G^\ast \subseteq \Gamma$. In general, some extremal particles survive the projection within $\Gamma - \Gamma_G^\ast$ when $\sigma_{0}^{(\phi)} \ne 1$, but all modes in $\Gamma - \Gamma_H^\ast$ are subextremal, hence the LWGC is violated when $H$ is non-trivial.

The argument for arbitrary Ricci flat manifolds is essentially the same, except that $\widehat{M}_{k-p}$ is generically no longer flat but only Ricci flat. In particular, by theorem 4.1 of~\cite{Fischer1975}, an arbitrary compact Ricci-flat manifold $M_k$ can be written as the finite quotient
\be \label{eqn:Ricciflatquotient}
M_k = \frac{T^p \times \widehat{M}_{k-p}}{G} \,,
\ee
where $G$ is a finite freely acting group which combines a translational subgroup of $T^p$ with a discrete isometry group of $\widehat{M}_{k-p}$, such that each nontrivial element of $G$ acts non-trivially on both factors and $\widehat{M}_{k-p}$ has no $G$-invariant Killing vectors. The latter implies that the Ricci-flat metric factorizes, as before.

By an identical argument to before, the finite coarseness sublattice $\Gamma_G^\ast \subseteq \Gamma$ is populated with particles saturating~(\ref{eqn:KKineq}). To establish that this matches the extremality bound, we first note that $\nu = 0$ KK modes for compactification on $T^p \times \widehat{M}_{k-p}$ are extremal, as only the pure gravity sector from compactification on $\widehat{M}_{k-p}$ contributes to $U(1)^p$ charged black holes upon toroidal compactification. Taking the $G$ quotient removes some of the massless fields, but none of the fields from this sector, hence the black hole solutions are unchanged, and the extremality bound is~(\ref{eqn:KKineq}).

\subsection{Gauge theories and Wilson lines}

So far we have discussed pure gravity theories compactified on Ricci flat manifolds and shown that their KK spectrum satisfies the sLWGC but not the LWGC or other previously conjectured strong forms of the WGC. The original motivation for a lattice conjecture in~\cite{Heidenreich:2015nta} was that gauge theories which satisfy the WGC can violate it upon toroidal compactification. By contrast, the LWGC, sLWGC, and similar lattice conjectures are robust under toroidal compactification.

In this section, we revisit this question in light of the LWGC counterexamples discussed above. We will argue that, while the LWGC is not preserved by KK reduction on general Ricci flat manifolds, the sLWGC is.

We begin by reviewing the toroidal case. A general two-derivative action for Abelian gauge fields coupled to moduli and gravity is
\be
S = \frac{1}{2\kappa_d^2} \int d^d x \sqrt{-g} \left[ \cc{R} - \frac{1}{2} h_{i j}(\phi) \nabla \phi^i \cdot \nabla \phi^j \right]-\frac{1}{2 e^2} \int d^d x \sqrt{-g}\, f_{a b}(\phi) F^a \cdot F^b \,,
\ee
where $h_{i j}$ is the metric on moduli space and $f_{a b}$ is the gauge kinetic matrix, both positive definite and in general $\phi$-dependent. Unfortunately, to our knowledge it is not possible to derive the black hole extremality bound in closed form for arbitrary functions $h_{i j}(\phi)$ and $f_{a b}(\phi)$.

Instead of attempting to solve the general problem, we focus on a few simple situations where the answer is known and consider the effects of compactification in these cases. The simplest of these is the Reissner-Nordstr\"om case, where the moduli are absent. Another simple case is that of a dilaton $\phi$ with a universal coupling $f_{a b}(\phi)= e^{-\alpha \phi} \hat{f}_{a b}$ to the gauge fields which are sourced by the black hole,\footnote{Here we fix a unit metric $h_{\phi \phi} = 1$ by a choice of coordinate on the one-dimensional scalar manifold.} where the $\alpha \to 0$ decoupling limit reproduces the Reissner-Nordstr\"om case. This ``Einstein-Maxwell-dilaton'' class of theories---which appears in many simple string theory examples---was considered in~\cite{Heidenreich:2015nta}, where it was shown that the extremality bound is unchanged after KK reduction on a circle. Here the coupling of the radion to the gauge field plays an essential role; if the radion acquires a mass, the extremality bound becomes stronger.

Although the KK reduced theory has two moduli, the dilaton and the radion, only one couples to the gauge field, and the other vanishes in the black hole background. Thus, the resulting theory is still in the Einstein-Maxwell-dilaton class, and we can consider further circle compactifications with the same result. Therefore, superextremal particles give rise to superextremal zero modes upon toroidal compactification, with the same mass and charge as in the parent theory.

This result can be extended to include black holes which are charged under the graviphotons on the compact torus, $T^k$. As explained in~\cite{Heidenreich:2015nta}, black hole solutions of this type can be obtained by boosting a black hole with no graviphoton charge along the torus. To be precise, the black hole solution lifts to a black brane solution with a smeared charge density in the higher dimensional theory. The black brane can be thought of as a sheet of charged particles spread out in the torus directions, and the result of boosting is analogous to the single-particle case. The extremality bound is therefore closely related to the KK mass formula (see e.g.~\cite{Heidenreich:2015nta}):
\be
M^2 \ge \gamma Q^2 + \frac{Q_{KK}^2}{R^2} \,,
\ee
where the constant $\gamma$ is that which sets the extremality bound in the absence of graviphoton charge.

Thus, the KK modes of superextremal particles are likewise superextremal. This guarantees that if a lattice conjecture is satisfied in the original theory (either the LWGC or the sLWGC will do, as will other variants) then it remains true in the KK reduced theory. It does \emph{not} ensure that the (original, non-lattice) WGC follows from its higher dimensional counterpart, however. For instance, the KK modes of any finite number of massive superextremal particles in the original theory will fail to satisfy the WGC in the KK reduced theory if $R$ is allowed to take arbitrarily small values.  This failure occurs for every value of $R$ if the original particles are exactly extremal. Therefore, while lattice WGC conjectures behave well under KK reduction, the WGC itself does not~\cite{Heidenreich:2015nta}.

Note that we are free to turn on Wilson lines $a^{a i} = \int A^a d \theta^i$ in the above examples. These appear in the extremality bound, for instance
\be
M^2 \ge \gamma Q^2 + \frac{1}{R^2} \biggl(Q_{KK}-\frac{a}{2 \pi} Q\biggr)^2
\ee
for compactification on a circle. However, $a$ has a shift symmetry in the low energy effective action, which forces it to appear in the same fashion in the extremality bound and KK mass formula. Thus, the KK modes remain superextremal with $a \ne 0$, despite changes in the charged spectrum.

We now consider compactification of Einstein-Maxwell-dilaton theory on a general Ricci flat manifold $M_k$. Due to the quotient structure~(\ref{eqn:Ricciflatquotient}), this is closely related to toroidal compactifications, except that $G$ will remove some of the KK modes of the superextremal charged particles. If these particles are scalars then $\sigma_0^{(\phi)} = 1$, and the KK modes on $\Gamma_G^{\ast}$ will be superextremal as before. Likewise, if they are bosonic fields of higher spin then polarizations exist for which $\sigma_0^{(\phi)} = 1$, and the same conclusion follows. However, for a fermion $\psi$ $\widehat{M}_{k-p}$ admits a zero mode if and only if it has a covariantly constant spinor, hence special holonomy $SU(n)$, $G_2$, $Spin(7)$, or a subgroup. If not, then superextremal fermions need not generate any superextremal KK modes, and the sLWGC can fail.

It should be noted, however, that our discussion implicitly relies upon supersymmetry, as we assume the presence of unstabilized moduli such as the radion. Unbroken supersymmetry requires not only that $\widehat{M}_{k-p}$ admits a covariantly constant spinor but also that $M_k$ does, equivalently that $\sigma_0^{(\psi)} = 1$ for at least one such spinor. In this case, the KK modes on $\Gamma_G^{\ast}$ will again be superextremal, and the sLWGC is preserved: if $\Gamma_1$ denotes a sublattice satisfying the conjecture before compactification, then $\Gamma_1\times \Gamma_G^{\ast}$ does so afterwards.

Besides the issue with special holonomy, so far the distinction between the sLWGC and the LWGC arises purely in the graviphoton sector, with the additional gauge fields coming along for the ride. This changes if we allow for \emph{discrete Wilson lines}, which can arise when $M_k$ has torsion one-cycles. Suppose $M_k$ has no continuous isometries, and let $\Sigma_r$ be a non-trivial torsion one cycle with $r \Sigma_r \cong 0$. This implies that
\be
\frac{1}{2\pi} \int_{\Sigma_r} A \in \bb{Z}/r \,,
\ee
since $\frac{1}{2\pi} \int A \mapsto \frac{1}{2\pi} \int A+1$ is a large gauge transformation.

To determine the effect of this Wilson line on the spectrum, it is useful to start with the $r$-fold cover $\widehat{M}_k$ (where $M_k=\widehat{M}_k/\bb{Z}_r$) on which $\Sigma_r$ is trivial. We then quotient by
\be
\frac{T^N \times \widehat{M}_k}{\bb{Z}_r} \,,
\ee
where $T^N$ represents the abstract torus associated to the $U(1)^N$ gauge group and the quotient combines gauge transformations (translations on $T^N$) with a discrete $\bb{Z}_r$ isometry of $\widehat{M}_k$. The situation is now very similar to the discussion in section~\ref{subsec:Ricciflat}. On the sublattice $\Gamma_{\bb{Z}_r}^\ast \subset \Gamma$ the $\nu = 0$ KK modes of fields whose zero modes survive the projection will be extremal. On the rest of the lattice $\Gamma - \Gamma_{\bb{Z}_r}^\ast$ there is no guarantee of superextremal modes, depending on $\sigma_{\nu}^{(\phi)}$, hence the sLWGC is preserved but not in general the LWGC.

In fact, the similarity between these two examples is no accident. On the one hand, although the abstract torus description of $U(1)^N$ applies more generally, this torus can be realized as an actual spatial torus in the case of Einstein-Maxwell-dilaton theory for a specific value of the dilaton coupling corresponding to an extra-dimensional radion. On the other, the $G$ quotient in~(\ref{eqn:Ricciflatquotient}) can be thought of as a set of discrete Wilson lines if we first compactify on $T^p$ and then compactify the resulting graviphoton gauge theory on $\widehat{M}_{k-p}$. The pure gravity example is therefore a special case of the more general phenomenon of discrete Wilson lines.

With this in mind, we give a brief, general discussion of discrete Wilson lines on manifolds without graviphotons; the case with graviphotons can be treated by first reducing on the torus $T^p \subset M_k$ and then treating them like the other gauge fields. Let $M_k$ be a Ricci flat manifold with no continuous isometries, and let $\widehat{M}_k$ be a finite cover with covering group $G$, $M_k = \widehat{M}_k/G$. We consider the quotient
\be
\frac{T^N\times \widehat{M}_k}{G}\,,
\ee
where $T^N$ is the abstract torus of the gauge group $U(1)^N$, as above, and $G$ acts freely on $T^N$. The $G$ quotient has the action
\be
g: \phi_{\vec{Q},\nu} \mapsto e^{2 \pi i \vec{g}\cdot \vec{Q}} \sigma_\nu^{(\phi)}(g) \phi_{\vec{Q},\nu} 
\ee
on the KK modes, where $\vec{Q}$ denotes the $U(1)^N$ charge, $\nu$ labels the $\widehat{M}_k$ mode expansion, $g \mapsto \vec{g}$ defines the $G$ action on $T^N \cong \bb{R}^N/\bb{Z}^N$, and $\sigma_\nu^{(\phi)}(g)$ is a mode and field-dependent phase factor.

Provided that $M_k$ has special holonomy, the argument for the sLWGC is largely parallel to before. For each superextremal field, there is some polarization with $\sigma_{0}^{(\phi)} = 1$, hence everywhere on the sublattice $\Gamma_G^\ast \subseteq \Gamma \equiv \bb{Z}^N$ there is at least one $\nu = 0$ mode which survives the $G$ projection. The mass and charge of this mode are equal to those of the parent field. Likewise, the $G$ projection does not affect the extremality bound because the charges are $G$ invariant, hence the corresponding black hole solutions must also be $G$ invariant, therefore modes which are projected out by $G$ are not turned on in these backgrounds.

Thus, the sLWGC is satisfied with a superextremal lattice $\Gamma_G^\ast$, but the LWGC can be violated.

\subsection{General considerations}

In the preceding discussion, we have presented evidence for the Sublattice Weak Gravity Conjecture using effective field theory, and in particular Kaluza-Klein theory. While our examples are not UV complete, every type of example we discuss above has specific realizations in supersymmetric string compactifications, including in particular the simple toroidal orbifold of section~\ref{subsec:simpleexample}.\footnote{In addition to the $T^6/(\bb{Z}_2\times\bb{Z}_2')$ realization discussed later, the $T^3/(\bb{Z}_2\times\bb{Z}_2')$ example of section~\ref{subsec:simpleexample} can be directly realized as a type IIA orientifold for which the involution has no fixed points, and which therefore lacks orientifold planes. Despite the different UV realizations, KK modes behave analogously in the two cases.} As we will see explicitly in the next section, the equivalent string theory examples include additional heavy charged particles, but the light charged spectrum matches the Kaluza Klein spectrum for large radius compactifications, hence many of the relevant features we have seen will necessarily apply to the corresponding string theory examples.

It may be possible to generalize our effective field theory discussion yet further to make contact with additional string theory examples. For instance, we could consider the dimensional reduction of higher dimensional theories with extended objects and $p$-form gauge fields $C_p$. This can be done without knowing much detail about the UV description of these objects provided that their fluctuations are unimportant to the problem at hand. This is true, for instance, when we compactify such a theory on a smooth manifold with $(p-1)$-cycles $\Sigma_{p-1}$ and study the lightest particles with a given charge under the gauge field arising from $C_p$ reduced on $\Sigma_{p-1}$, which correspond to branes wrapped on $\Sigma_{p-1}$. So long as the probe approximation is valid, the mass of the charged particle is simply the tension of the brane times the volume of the cycle, and one can in principle check whether a spectrum of branes satisfying the WGC leads to superextremal particles in the lower dimensional theory. It may even be possible to check lattice statements in this context.

However, computing the volumes of arbitrary cycles on a manifold of special holonomy such as a Calabi-Yau manifold is a very difficult problem. We will not attempt it here, and a realistic approach to such a problem seems to be outside presently available theoretical tools, except perhaps for numerical methods. A simpler, related problem in Calabi-Yau geometry is the counting of holomorphic curves.  These curves are calibrated with respect to the K\"ahler form, enabling the computation of their volumes even without complete knowledge of the Calabi-Yau metric.  Applied to M-theory compactification on a Calabi-Yau three-fold, the sLWGC implies the existence of a holomorphic curve in every effective two-cycle class within a finite index sublattice of the homology lattice $H_2$.  For further discussion, see \cite{Heidenreich:2016jrl}.

One could also try to realize some analog of string theory flux compactifications in an effective field theory description. However, this is very difficult due to well-known no-go theorems such as~\cite{Maldacena:2000mw}, which require a positive scalar potential, higher derivative corrections, or negative tension branes in order to turn on fluxes in a compactification with vanishing cosmological constant.\footnote{Note, however, that scalar gradients are allowed (as are $F_{d-1}$ fluxes dual to axion gradients). It might be interesting to explore this possibility further.} In the first two cases, this requires introducing a UV scale, which calls into question the validity of the effective field theory description, whereas the last requires careful handling to avoid violating unitarity. All three may be realized in string theory, but a purely effective field theory description is difficult without putting in substantial UV information.

We now move on to discuss the role of the sLWGC in string theory.

\section{The sLWGC and String Theory}

We have seen that general Kaluza-Klein compactifications motivate the sLWGC. Here we will explore its validity in perturbative string theory. We will start by giving a very general argument that connects the sLWGC to modular invariance, before presenting concrete examples that illustrate how the full LWGC can fail.

\subsection{The sLWGC from modular invariance} \label{subsec:modularinvariance}

In the context of perturbative string theory, the sLWGC is related to modular invariance. For every NSNS sector gauge field, there is a corresponding conserved current in the worldsheet CFT.
 By introducing a chemical potential for this current we can track the charges of the states in the perturbative string theory spectrum. Imposing modular invariance, we demonstrate that this spectrum satisfies the sLWGC, up to a few technical assumptions that we explain as they arise.

Note that RR gauge fields do not appear as conserved currents in the worldsheet CFT because the charged states, coming from D-branes, are nonperturbatively heavy in string units in the weak coupling limit. Since they do not appear in the perturbative spectrum, nothing is charged, and there is no conserved current. The same considerations apply to the magnetic duals of NSNS sector gauge fields, where the charged states come from, e.g., NS5 branes. Likewise, in type I string theory, though there are charged states in the spectrum, the massive states which satisfy the sLWGC are nonperturbatively heavy, and correspond to the excitations of fundamental strings in the heterotic S-dual. Consequently, the following argument applies only to electric NSNS sector gauge fields in closed string theory.

The worldsheet CFT has a partition function of the form:
\be
Z (\mu, \bar{\mu} ; \tau, \bar{\tau}) \equiv \Tr (q^{\Delta} 
   \bar{q}^{\tilde{\Delta}} y^Q  \bar{y}^{\tilde{Q}}) \,,
\ee
where $\Delta = L_0 - \frac{c}{24}$, $\tilde{\Delta} = \tilde{L}_0 -
\frac{\tilde{c}}{24}$, $Q$ and $\tilde{Q}$ are the charges carried by
left/right movers under a conserved current, $q = e^{2 \pi i \tau}$ and $y = e^{2 \pi i \mu}$. We have
\begin{equation} \label{eqn:rhoperiod}
  Z (\mu + \rho) = Z (\mu) \;, \qquad \forall \rho \in \Gamma_Q^{\ast} \,,
\end{equation}
where $\Gamma_Q^{\ast} = \{ (\rho, \tilde{\rho}) | \rho Q - \tilde{\rho}
\tilde{Q} \in \bb{Z} \}$ is the dual lattice to the charge lattice.
Following e.g.~\cite{Benjamin:2016fhe}, modular transformations act as
\begin{align}
  Z (\mu ; \tau + 1) = & Z (\mu ; \tau) \,, &
  Z (\mu / \tau ; - 1 / \tau) = & e^{\pi i k \frac{\mu^2}{\tau} - \pi i
  \tilde{k}  \frac{\bar{\mu}^2}{\bar{\tau}}} Z (\mu ; \tau) \,. \label{eqn:SL2Z}
\end{align}
Here $k, \tilde{k}$ are related to the leading term in the current-current
OPE:
\begin{equation}
  J_L (z) J_L (0) \sim \frac{k}{z^2} + \ldots \;, \qquad J_R (\bar{z}) J_R (0)
  \sim \frac{\tilde{k}}{\bar{z}^2} + \ldots \;.
\end{equation}
Unitarity requires that $k, \tilde{k}$ are
non-negative, and positive for non-trivial currents.

In general, we should consider $\mu$ and $\bar{\mu}$ to be independent variables because there are two separate conserved currents $J_L$ and $J_R$ and their corresponding charges $Q$ and $\tilde{Q}$. For example, a periodic free boson has two conserved charges, one left-moving and one right-moving:
\begin{equation}
  Q = \sqrt{\frac{\alpha'}{2}} \biggl[\frac{n}{R} + \frac{w R}{\alpha'}\biggr] \qquad, \qquad \tilde{Q} = \sqrt{\frac{\alpha'}{2}}\biggl[\frac{n}{R} - \frac{w R}{\alpha'}\biggr] \,,
\end{equation}
where $n, w \in \bb{Z}$ are the momentum and winding quantum numbers. The corresponding string theory has two gauge bosons coming from the off-diagonal components of the metric and B-field.

Moreover, in examples such as heterotic string theory, the number of left-moving and right-moving currents is usually different, hence there is no way to identify the chemical potentials in complex conjugate pairs. To avoid this notational problem, we write $\tilde{\mu}$ henceforward in place of $\bar{\mu}$, and use the shorthand
\begin{align}
y^Q &\equiv \exp[ 2 \pi i \mu_a Q^a] \,, & \tilde{y}^Q &\equiv \exp[- 2 \pi i \tilde{\mu}_a \tilde{Q}^a] \,,
\end{align}
for the case of multiple currents, where now
\begin{equation}
  J_L^a (z) J_L^b (0) \sim \frac{k^{a b}}{z^2} + \ldots \;, \qquad J_R^{\tilde{a}} (\bar{z}) J_R^{\tilde{b}} (0)
  \sim \frac{\tilde{k}^{\tilde{a} \tilde{b}}}{\bar{z}^2} + \ldots \;.
\end{equation}

Combining the periodicity condition~(\ref{eqn:rhoperiod}) with the S-duality transformation~(\ref{eqn:SL2Z}), we
obtain:
\begin{equation} \label{eqn:Zquasiperiod}
  Z (\mu + \tau \rho ; \tau) = \exp\biggl[- 2 \pi i k \mu \rho - \pi i k \rho^2 \tau +
  2 \pi i \tilde{k}  \tilde{\mu}  \tilde{\rho} + \pi i \tilde{k}  \tilde{\rho}^2
  \tau\biggr] Z (\mu ; \tau) \,.
\end{equation}
In the case of multiple currents, we have $k \mu \rho \mapsto k^{a b} \mu_a \rho_b$, etc.. For simplicity, we suppress $k^{a b}$ and $\tilde{k}^{a b}$ henceforward, treating them as metrics which appear as needed depending on index positions. Thus, for instance $\mu \cdot \rho \equiv \mu_a k^{a b}
\rho_b$, $\mu \cdot Q \equiv \mu_a Q^a$, and $\tilde{Q}^2 \equiv \tilde{Q}^{\tilde{a}} \tilde{k}^{- 1}_{\tilde{a} \tilde{b}} \tilde{Q}^{\tilde{b}}$.

Since the partition function enumerates the spectrum of the theory, the
quasi-period $\mu \rightarrow \mu + \tau \rho$ must map the spectrum to
itself. To describe this, we define:
\begin{equation}
  T \equiv \Delta - \frac{1}{2} Q^2 \qquad, \qquad \tilde{T} \equiv
  \tilde{\Delta} - \frac{1}{2}  \tilde{Q}^2 \,.
\end{equation}
The condition~(\ref{eqn:Zquasiperiod}) can now be written as:
\begin{equation}
  Z = \Tr \left( q^{T + \frac{1}{2} Q^2}  \bar{q}^{\tilde{T} + \frac{1}{2} 
  \tilde{Q}^2} y^Q  \tilde{y}^{\tilde{Q}} \right) = \Tr \left( q^{T +
  \frac{1}{2}  (Q + \rho)^2}  \bar{q}^{\tilde{T} + \frac{1}{2}  (\tilde{Q} +
  \tilde{\rho})^2} y^{Q + \rho}  \tilde{y}^{\tilde{Q} + \tilde{\rho}} \right) \; .
\end{equation}
(Note that $(Q + \rho)^a = Q^a + k^{a b} \rho_b$ contains implicit
$k$-dependence.) By expanding the trace in powers of $Q$, and matching the two
sides, we see that the spectrum must be invariant under
\begin{equation}
  Q \rightarrow Q + \rho \qquad, \qquad \tilde{Q} \rightarrow \tilde{Q} +
  \tilde{\rho} \,,
\end{equation}
with $T$ and $\tilde{T}$ held fixed. The rearrangement of the spectrum under simultaneous changes in charge and conformal weight may be familiar as spectral flow \cite{Schwimmer:1986mf}. Notice that this invariance implies
\begin{equation}
  \Gamma_Q^{\ast} \subseteq \Gamma_Q \,.
\end{equation}
This is a modular invariance condition similar to (but weaker than) the
self-duality condition for Narain compactification.\footnote{A self-dual
charge lattice $\Gamma_Q^{\ast} = \Gamma_Q$ would imply the LWGC, as in,
e.g., toroidal compactifications of the heterotic string.} The quotient space $\Gamma_Q/\Gamma_Q^{\ast}$ is a discrete group. In fact, although the charges and the level $k$ for an abelian group can always be rescaled, if we choose a basis in which the charge lattice is simply $\bb{Z}^p$ the level of the gauge group acquires physical meaning: it determines the index of the sublattice $\Gamma_Q^\ast$ within $\Gamma_Q$.

Using this constraint, we now argue that for any $Q \in \Gamma_Q^{\ast}$,
there exists a level-matched state of charge $Q$ and mass
\begin{equation} \label{eqn:worldsheetWGC}
  \Delta = \tilde{\Delta} = \frac{\alpha'}{4} m^2 \leqslant \max \left(
  \frac{1}{2} Q^2, \frac{1}{2}  \tilde{Q}^2 \right) \,.
\end{equation}
To do so, we note that there is a graviton in the spectrum, with $\Delta =
\tilde{\Delta} = 0$ and $Q = \tilde{Q} = 0$, hence $T = \tilde{T} = 0$.
Applying the above transformation, we obtain states with charges $Q = \rho$,
$\tilde{Q} = \tilde{\rho}$ and
\begin{equation}
  \Delta = \frac{1}{2} Q^2 \qquad, \qquad \tilde{\Delta} = \frac{1}{2} 
  \tilde{Q}^2 \,.
\end{equation}
Since $\Delta - \tilde{\Delta} \in \bb{Z}$ is required for modular
invariance (hence $\Gamma_Q^{\ast}$ must be an even lattice), we can turn on integer-moded oscillators in one of the non-compact dimensions to match levels, $\Delta = \tilde{\Delta}$. Since the oscillators for two dimensions are removed by gauge fixing, this is possible for $D \geqslant 3$ non-compact directions in the target space, and the result saturates~(\ref{eqn:worldsheetWGC}).

Conversely, because the CFT spectrum must have a bottom, there is a lower bound on the masses of level-matched states at large charge. In particular, requiring $\Delta \geqslant
\Delta_{\min}$ and $\tilde{\Delta} \geqslant \tilde{\Delta}_{\min}$ for some $\Delta_{\min}$ and $\tilde{\Delta}_{\min}$ implies that $T \geqslant T_{\min}$
and $\tilde{T} \geqslant \tilde{T}_{\min}$ for some $T_{\min}$ and $\tilde{T}_{\min}$. Thus, all level-matched states in
the spectrum satisfy:
\begin{equation}
  \frac{\alpha'}{4} m^2 \geqslant \max \left[ T_{\min} + \frac{1}{2} Q^2,
  \tilde{T}_{\min} + \frac{1}{2}  \tilde{Q}^2 \right] \,.
\end{equation}
Asymptotically at large charge this implies
\begin{equation} \label{eqn:asymptotic-lower-bound}
  m \gtrsim \sqrt{\frac{2}{\alpha'}} \max \biggl( \sqrt{Q^2}, \sqrt{\tilde{Q}^2} \biggr) +\mathcal{O} \left( \frac{1}{Q} \right) \,,
\end{equation}
which matches~(\ref{eqn:worldsheetWGC}) at leading order in large charge.

Combining~(\ref{eqn:worldsheetWGC}) and~(\ref{eqn:asymptotic-lower-bound}), we conclude that the sLWGC holds for the $g_s \to 0$ spectrum if and only if the black hole extremality bound
is \emph{at least as strong as}:
\begin{equation} \label{eqn:candidateExtremality}
  \frac{\alpha'}{4} m^2 \geqslant \max \left( \frac{1}{2} Q^2, \frac{1}{2} 
  \tilde{Q}^2 \right) \,.
\end{equation}
This is rather difficult to show in general, as it requires us to understand the effective action and the possible black hole solutions to it.
Since highly excited strings turn into black holes~\cite{Horowitz:1996nw,Horowitz:1997jc}, we might be inclined to match the large charge asymptotics of the perturbative string spectrum with the black hole spectrum, which would imply that~(\ref{eqn:candidateExtremality}) matches the extremality bound. In directions with BPS states, this is obviously correct, since the light BPS particles and heavy BPS extremal black holes have the same charge-to-mass ratio. Moreover, we give an argument below that this matching holds for most, if not all, orbifolds. However, this kind of reasoning is flawed in general because the perturbative worldsheet description ceases to apply once the states transition to black holes, and we have no direct control over the slope of the spectrum above this transition based solely on the slope below it.

To establish that~(\ref{eqn:candidateExtremality}) is the extremality bound for a large class of orbifolds, we observe that the naive connection between the asymptotic slope of the perturbative spectrum and the extremality bound holds in toroidal compactification of both type II and heterotic string theories. In the former case, this follows from BPS bounds, whereas in the latter it has been checked explicitly in the literature~\cite{Sen:1994eb}. Orbifolding by a discrete group $G$ removes some of the fields, but the gauge fields which survive are $G$ invariant, hence the corresponding black hole solutions are also $G$ invariant, and none of the projected out fields is turned on in the original background. This implies that the extremality bound remains the same as in the parent theory, with one possible caveat: the orbifold introduces twisted sectors, potentially including additional massless particles. If these include scalars which develop a tadpole in the black hole background in question, then the black hole solution will be modified. However, this can be ruled out in many examples because the twisted sectors are often charged under the gauge group, and charged fields cannot acquire tadpoles without Higgsing the gauge group. Even in examples where the twisted sector is neutral, it seems unlikely that any of these fields will be turned on.

Conversely, it is also possible that additional massless gauge bosons could appear in the twisted sectors. This can happen in type II string theory, but typically only for RR gauge bosons; non-trivial examples may exist in heterotic string theory, though none that we have constructed. If this occurs, the above argument cannot be used to show that (\ref{eqn:candidateExtremality}) is the correct extremality bound for black holes carrying these charges. We are then left with the same unclear situation as in more general worldsheet theories.

More generally, if a well controlled example can be constructed for which the extremality bound is weaker than~(\ref{eqn:candidateExtremality}) in any direction in charge space, this would give a counterexample to the sLWGC. We leave further study of this point to future work.

One other subtlety is that, in the presence of spacetime fermions, the modular
invariant combination is
\begin{equation}
  Z (\mu, \bar{\mu} ; \tau, \bar{\tau}) \equiv \Tr [(- 1)^F q^{\Delta} 
  \bar{q}^{\tilde{\Delta}} y^Q  \bar{y}^{\tilde{Q}}] \,,
\end{equation}
where $F$ is the spacetime fermion number. If the background is
supersymmetric, then the partition function vanishes identically. In
particular, the graviton contribution is cancelled by those of its
superpartners, and the above argument is technically incorrect. However, this
technical objection should have a technical answer. The modular
invariance condition of one loop amplitudes with insertions is non-trivial,
and (as is usually assumed without comment) should imply naive arguments
which ignore these cancellations. One way to address this problem is to turn on a chemical potential for spacetime rotations, which can split bosons from fermions. We sketch the details of this approach in Appendix \ref{app:modinvrot}.

\subsection{Examples}\label{ssec:examples}

In this section, we describe a handful of theories arising from toroidal orbifold compactifications of type II and heterotic string theory. As implied by the modular invariance argument given above, the sLWGC is indeed satisfied in all of these examples, which can be verified on a case by case basis. However, these examples violate the other strong forms of the WGC discussed previously, such as the LWGC and the strong form of~\cite{Brown:2015iha}.

To begin, we consider type II string theory on the $T^6/(\bb{Z}_2 \times \bb{Z}_2')$ orbifold 
 with orbifold action defined by the two generators
\begin{align}
\begin{split}
\theta: \quad & \theta_4 \mapsto \theta_4 + \pi,\, \theta_5 \mapsto \theta_5 + \pi, \\
\omega: \quad &  \theta_6 \mapsto \theta_6 + \pi, \, \theta_i \mapsto -\theta_i, i=1,...,4.
\end{split}
\label{eq:favorite}
\end{align}
For simplicity, we take the metric to be diagonal in the $\theta_I$ basis, $I=1,\ldots,6$.  This orbifold preserves sixteen real supercharges in four dimensions.  Note that the orbifold has no fixed points so the resulting manifold is flat; as a result, this example can be understood in the supergravity description as well as on the worldsheet.  Indeed, the physics of this example is very similar to that of the effective theory described in section \ref{subsec:simpleexample}, and much of the analysis carries over straightforwardly via the dictionary $\theta_w\rightarrow \theta_4, \theta_y \rightarrow \theta_5, \theta_z \rightarrow \theta_6$.  Any particle of odd KK charge around $\theta_5$ and no other gauge gauge charge will be subextremal, as the orbifold projection forces an additional contribution $(n_4/R_4)^2$ to its mass squared:
\begin{equation}
 m^2 = \left(\frac{n_5}{R_5} \right)^2 + \left(\frac{n_4 }{R_4} \right)^2\,,~~n_5~ \rm{ odd },\, n_4~ \rm{ odd.}
\label{eq:untwistedmass}
\end{equation}
Here, it is crucial that $n_4$ does not correspond to a gauge charge, as the orbifold action projects out the gauge boson associated with KK charge around $\theta_4$.  However, $n_5$ does correspond to a conserved gauge charge, which makes any decay of this particle to massless uncharged particles kinematically impossible.

One distinction between the UV complete theory and the Kaluza-Klein effective description discussed before is the presence of twisted sectors. Naively, this difference could affect conclusions about superextremal particles far out on the charge lattice.
For example, the $\omega$-twisted sector states of odd KK charge around $\theta_5$ are BPS and therefore superextremal. However, these necessarily have half-integer winding charge around $\theta_6$, so their charge differs from that of the untwisted sector states. Similar considerations apply to the other twisted sectors, implying that there are no superextremal particles with odd $n_5$ and integer (or vanishing) winding charges.

Clearly this violates the LWGC.
Depending on the radii $R_I$, additional ``strong forms" of the WGC can also be violated.\footnote{Since the radions are orbifold invariant and have no potential (due to the sixteen unbroken supercharges) we are free to choose any values for the $R_I$.} For example, taking $R_4 \gg R_5 > R_6$ and $R_I \gg \sqrt{\alpha'}$, the winding modes become heavy and the lightest charged particle in the spectrum is subextremal with $(n_5,n_6)=(1,0)$. This directly contradicts the strong form proposed in \cite{ArkaniHamed:2006dz}, and by extension that proposed in \cite{Brown:2015iha}.
This particle is also the state of smallest charge in its direction in the lattice, violating the other strong form discussed (and rejected) in \cite{ArkaniHamed:2006dz}. Thus, almost every WGC strong form that has been previously discussed in the literature is violated by this example, but notably the sLWGC is satisfied.

In fact, this theory has an even stranger property, which calls into question the original motivation behind the WGC: at tree level in string perturbation theory, all of the lightest particles with odd $n_5=P$ and no other gauge charge are stable. 
To see this, 
we first consider a two-body decay, and call their $n_5$ charges $p$ and $q= P -p$, respectively.  Without loss of generality, we may take $p$ to be odd and $q$ to be even.  A priori, these particles could carry charge under additional $U(1)$s, but given our assumptions about the radii the lightest particles with odd $n_5$ charge are those with mass $m^2 = n_5^2/R_5^2+1/R_4^2$ and the lightest particles of even $n_5$ charge are BPS states of mass $m=n_5/R_5$.  Thus, it suffices to consider only these final states to check if any decay is kinematically allowed. This would require
\begin{equation}
\sqrt{\biggl(\frac{p+q}{R_5} \biggr)^2 + \biggl(\frac{1}{R_4} \biggr)^2} \ge  \sqrt{ \biggl(\frac{p}{R_5} \biggr)^2 + \biggl(\frac{1}{R_4} \biggr)^2 } + \left|\frac{q}{R_5} \right|
\end{equation}
for some $q\ne 0$, but this fails by the triangle inequality. This argument is easily extended to an arbitrarily large number of particles in the final state: upon replacing any pair of particles in the final state by the lightest single-particle state with the same $n_5$ charge the mass is non-increasing.\footnote{When the two particles have charges of opposite parity, this follows from the above argument. When they have the same parity, this follows from the BPS bound.} Doing this repeatedly shows that the final state is necessarily heavier than the initial state, and the decay is kinematically forbidden.

Thus, if we take the tree-level string spectrum at face value, we conclude that there is an infinite tower of stable particles with odd $n_5$ charge. For large charges, these particles become black holes. Since they are not BPS, these black holes are both (very slightly) subextremal and stable, seemingly in violation of the spirit of the WGC. Indeed, this example highlights the fact that the existence of a superextremal particle is insufficient to guarantee that every charged black hole can decay~\cite{Bachlechner}.

However, the tree level calculation breaks down before the black hole transition occurs~\cite{Horowitz:1996nw,Horowitz:1997jc}, so the above conclusions do not follow. We can instead constrain what happens above this threshold using the BPS bound. There are three possibilities:\footnote{Here we only consider objects with $n_5$ charge and vanishing charge under the other $U(1)$s. Allowing for other charges does not change the outcome in any fundamental way.} 
\begin{enumerate}
\item
There are BPS black holes with odd charge.
\item
For fixed $g_s$, there exists $\Delta>0$ such that $M-M_{\rm BPS}\ge \Delta$ for all states with odd charge.
\item
There is an infinite tower of stable subextremal states with odd charge.
\end{enumerate}
In the first case, any given BPS state with odd charge must disappear for $g_s < g_c$ for some critical coupling $g_c>0$, since these states are absent from the perturbative string spectrum. (This could happen via the usual pairing of several BPS states into a long multiplet.) If there is an odd-charged BPS state in the spectrum at every value of $g_s$ then there must be an infinite number of such critical couplings, accumulating near $g_s = 0$. This sounds implausible, but not impossible.

In the second case, the spectrum of large semiclassical black holes is modified in an unexpected way: if the black hole has odd charge, then the extremality bound is slightly stronger, by a fractional amount $\Delta / M = (\Delta/M_{\rm Pl}) (M_{\rm Pl}/M)$. This effect is competitive with curvature corrections (which, however, do not modify the extremality bound due to the BPS condition), but it is unclear how to generate it by corrections to the effective action, which do not obviously depend on the parity of the black hole charge.

If we reject the first two possibilities then the third follows as we now argue. An odd charged black hole $i$ can decay if a lighter odd-charged black hole or particle $j$ exists with $\Delta_j \le \Delta_i$, $\Delta_i \equiv M_i - M_{\rm BPS}$. Since the spectrum of $\Delta$s is positive but without a positive lower bound, scanning the odd-charged black hole spectrum from small to large charge we necessarily find an infinite number of black holes with $\Delta$ less than all previous entries, hence there is an infinite tower of stable subextremal states.

\medskip

For a higher-dimensional example of a theory violating the LWGC, we consider an $E_8 \times E_8$ heterotic compactification on $T^4/\bb{Z}_2$ with Wilson lines.  Such a compactification preserves eight real supercharges in 6d.  We use the same setup as the above type II example but take $\theta_5$ and $\theta_6$ to be directions along the internal even-self-dual lattice $\Gamma_{16}$:\footnote{In our conventions, either $\Gamma_{16} = \Gamma_8^+ \otimes \Gamma_8^+$ or $\Gamma_{16}=\Gamma_{16}^+$, with $\Gamma_k^+ \equiv \Gamma_k^0 \cup \left[\Gamma_k^0+(\frac{1}{2},\ldots,\frac{1}{2})\right]$ where $\Gamma_k^0 \equiv \{(a_1,\ldots,a_k)|a_i \in\mathbb{Z}, \sum_i a_i \in 2\mathbb{Z}\}$ is the $SO(2k)$ root lattice.\label{fn:lattice}}
\begin{align}
\begin{split}
\theta: \quad & \theta_4 \mapsto \theta_4 + \pi,\, \Gamma_{16} \mapsto \Gamma_{16} + e_5, \\
\omega: \quad &  \theta_i \mapsto -\theta_i, i=1,...,4,\, \Gamma_{16} \mapsto \Gamma_{16} + e_6 \,,
\end{split}
\end{align}
where $e_5$ and $e_6$ are internal translation vectors with $2 e_{5,6} \in \Gamma_{16}$. These are restricted by modular invariance to satisfy $e_5^2 \in \bb{Z}$, $e_6^2 \in \bb{Z}+1/2$, $(e_5+e_6)^2 \in \bb{Z}+1/2$. We can choose for instance:
\begin{align}
e_5 &= (1,0,0,\ldots)\,, & e_6 &= \frac{1}{2} (1,-1,0,\ldots)\,,
\end{align}
in accordance with these requirements.

We will see that the LWGC is violated in the $\omega$-untwisted sector. To show this, we briefly review the spectrum.
The mass formula for the heterotic string in an untwisted sector is
\begin{equation} \label{eqn:levelmatching}
\frac{\alpha'}{4} m^2 = \frac{1}{2} (P^2+p_L^2) + N_L - 1 = \frac{1}{2} (r^2+p_R^2) + N_R - \frac{1}{2} \,.
\end{equation}
Here $P$ denotes the momentum along the internal lattice and $r$ is an $SO(8)$ weight constrained by the GSO projection\footnote{In particular, the GSO projection requires $r \in \Gamma_4^+ + (1,0,0,0)$ in the notation of footnote~\ref{fn:lattice}.} which arises from bosonizing the usual right-moving RNS worldsheet fermion.
In addition, $p_{L,R}$ are the spacetime momenta associated to the left and right and right movers and $N_{L,R} \in \mathbb{Z}_{\ge 0}$ are oscillator numbers.

The left and right-moving momenta on the fourth circle of the $T^4$ are
\begin{align}
p_L &= \sqrt{\frac{\alpha'}{2}} \biggl( \frac{n_4}{R_4} + \frac{w_4 R_4}{\alpha'} \biggr)\,, & p_R &= \sqrt{\frac{\alpha'}{2}} \biggl( \frac{n_4}{R_4} - \frac{w_4 R_4}{\alpha'} \biggr)\,,
\end{align}
where $n_4$ and $w_4$ are the momentum and winding quantum numbers,\footnote{In our conventions $R_4$ is the radius of the circle \emph{after} taking the $\theta_4$ quotient, and $n_4$ and $w_4$ are the associated momentum and winding quantum numbers.} fixed by the quantization conditions
\begin{align}
w_4 &\in \mathbb{Z}\,,&
P &\in \Gamma_{16}+w_4 e_5\,,&
n_4 &\in \mathbb{Z}-P\cdot e_5+w_4 e_5^2/2\,. \label{eqn:quantconds}
\end{align}
Let us consider a particle of charge $P = (1,0,0,\ldots)$, which requires $w_4 \in 2 \mathbb{Z}+1$ and $n_4 \in \mathbb{Z}+1/2$. The GSO projection fixes $r^2\ge 1$, so we find
\begin{equation}
\frac{\alpha'}{4} m^2 = \frac{1}{2} p_L^2 + N_L - \frac{1}{2} = \frac{1}{2} p_R^2 + N_R \le \frac{1}{2} 
\end{equation}
for a superextremal particle, which implies $N_R = 0$ as well as $N_L = 0,1$. One can check that the case $N_L = 1$ has a solution if and only if $R_4 = \sqrt{\alpha'/2}$, so we focus on the case $N_L=0$, which requires $w_4 = 2 n_4 = \pm 1$. We find a solution if and only if
\begin{equation}
\left(\frac{x-1/x}{2}\right)^2 \le 1 \mbox{\;with\;} x \equiv \sqrt{\frac{2}{\alpha'}} R_4 \,,
\end{equation}
hence if
\begin{equation}
(\sqrt{2}-1) \sqrt{\frac{\alpha'}{2}} \le R_4 \le (\sqrt{2}+1) \sqrt{\frac{\alpha'}{2}} .
\end{equation}
Outside this range of radii the LWGC is violated.

As before, this violation comes from the extra contribution of the spacetime momentum to the particle mass, which has no corresponding charge because the corresponding photons are removed by the $\mathbb{Z}_2$ orbifold. Unlike before, there is a tachyonic contribution to the mass coming from the zero-point energy of the string. As a consequence of this and of the purely left-moving gauge charges, farther out on the charge lattice there are superextremal particles of every charge. Suppose that $R_4 > \sqrt{\alpha'/2}$ and consider sites with
\be
P^2 \ge \frac{2 R_4^2}{\alpha'} \,.
\ee
For fixed $P$, the quantization conditions~(\ref{eqn:quantconds}) fix the parities of the integers $w_4$ and $2 k_4$, and there is a superextremal solution to the level-matching conditions~(\ref{eqn:levelmatching}) with $r^2 = 1$ and $N_L=0,1$ so long as we can solve
\be
p_L^2 \le 2\,, \qquad p_R^2 \le P^2 \,,
\ee
subject to the parity constraints. One such solution is:
\be
w_4 = -1\,, \qquad n_4 = \frac{R_4^2}{\alpha'} - \delta\,,
\ee
where $0 \le \delta < 1$ is chosen to satisfy $2 n_4 \in \mathbb{Z}$ along with the parity constraint on $n_4$.  Analogous reasoning applies to the case $R_4 < \sqrt{\alpha'/2}$ with $P^2 \ge \frac{\alpha'}{2 R_4^2}$ .
Thus, lattice sites with sufficiently large charge have superextremal particles, though the finite region where LWGC violations occur grows as $R_4$ is taken farther from the string scale in either direction.

Note that in this example the sLWGC is satisfied with coarseness 2, because if $P$ lies on the charge lattice then $2 P$ lies within the original charge lattice $\Gamma_{16}$ and moreover $(2 P)\cdot e_5 \in \mathbb{Z}$, hence $w_4$ and $2 n_4$ are even and we can set $p_L = p_R = 0$. This is in agreement with the modular invariance argument presented in the previous subsection.

While in this example the LWGC violation is limited to a finite number of lattice sites (for fixed $R_4$), KK reduction of this theory on a circle produces a 5d theory which violates the LWGC at an infinite set of points on the charge lattice.  In particular, all of the KK modes of the subextremal particles of the 6d theory will be subextremal in the 5d theory.  The sLWGC, on the other hand, is preserved under circle compactification.

We have explicitly verified that the sLWGC is satisfied in a large class of heterotic orbifold compactifications.  In particular, any supersymmetric $T^n/G, n\leq 6$ Abelian orbifold of the $E_8 \times E_8$ or $Spin(32)/\bb{Z}_2$ heterotic string with commuting Wilson lines and orbifold shifts along the internal left-moving $T^{16}$ will satisfy the sLWGC at tree level. In the absence of non-trivial space groups or Wilson lines, these theories  satisfy the full LWGC.

It is interesting to ask about how badly the LWGC can be violated in orbifold examples. We focus here on orbifolds of the heterotic string preserving eight supercharges, and distinguish two measures of failure: the coarseness and index of the sublattice $\Gamma_{\rm ext} \subseteq \Gamma$.  The largest coarseness we have found for such a sublattice is 3, coming from a $T^4/\mathbb{Z}_3$ heterotic orbifold with Wilson lines.  The largest index we have found is $2^8 = 256$, coming from a $T^4/\mathbb{Z}_2 $ heterotic orbifold with Wilson lines.  As an example of the former, consider the $T^4/\mathbb{Z}_3$ $E_8 \times E_8$ heterotic orbifold with orbifold action
\begin{align}
\begin{split}
\theta: \quad & z_1 \mapsto  e^{2 \pi i/3} z_1 ,\, z_2 \mapsto  e^{-2 \pi i /3} z_2 ,\,  \Gamma_{16} \mapsto \Gamma_{16} + e_0.
\end{split}
\end{align}
Here, $z_1$ and $z_2$ are complex numbers, each parametrizing a $T^2$ with complex structure $\tau = e^{ \pi i/3}$.  $e_0$ is an element of $\Gamma_{16}/3$, which we take to be
\begin{equation}
e_0 = \frac{1}{3} (1,-1,0,0,...,0).
\end{equation}
In addition, we turn on a Wilson line on the first torus,
\begin{equation}
e_1 = \frac{1}{3} (0,0,1,1,2,0,...,0).
\end{equation}
This combination solves the constraints imposed by modular invariance, $e_0^2 \in 2\mathbb{Z}/3  + 2/9$, $e_1^2 \in 2 \mathbb{Z}/3$, $e_0 \cdot e_1 \in \mathbb{Z}/3 $.  At small radius, one can check that untwisted sector states winding once around the first torus are subextremal for small values of the radius, similar to the previous $E_8 \times E_8$ heterotic orbifold example.  These states all lie in the shifted lattice $\Gamma_{16} + e_1$.  On the other hand, all of the charge sites in the $E_8 \times E_8$ lattice $\Gamma_{16}$ have superextremal states, and since $ 3 e_1 \in \Gamma_{16}$, we see that $\Gamma_{\rm ext}$ has coarseness 3.

For a $T^4/\mathbb{Z}_3$ orbifold, only two independent Wilson lines are allowed, since the Wilson lines around the A and B cycles of each $T^2$ are related by the orbifold action.  On the other hand, a $T^4/\mathbb{Z}_2$ orbifold allows independent Wilson lines for each $S^1$.  We can produce an example of a sublattice with an index as large as 256 by considering heterotic string theory compactified on a $T^4/\mathbb{Z}_2 \times S^1$ orbifold, turning on independent Wilson lines in $\Gamma_{16}/2$ around each of the $T^4$ directions twisted by $\mathbb{Z}_2$.  For instance, we may choose a shift in $\Gamma_{16}/2$ for the orbifold twist $\omega$:
\begin{equation}
e_0 = \frac{1}{2}(1,-1,0,...,0),
\end{equation}
and turn on Wilson lines for $\theta_i$, $i=1,2,3,4$:
\begin{equation}
\begin{split}
e_1 = \frac{1}{2} (\vec{1}_4,\vec{0}_4,\vec{0}_4,\vec{0}_4) \,,~ e_2 = \frac{1}{2} (\vec{0}_4,\vec{1}_4,\vec{0}_4,\vec{0}_4) \, \\ 
e_3 =\frac{1}{2} (\vec{0}_4,\vec{0}_4,\vec{1}_4,\vec{0}_4) \,,~ e_4 = \frac{1}{2} (\vec{0}_4,\vec{0}_4,\vec{0}_4,\vec{1}_4).
\end{split}
\end{equation}
Here, $\vec{1}_4 = (1,1,1,1)$ and $\vec{0}_4=(0,0,0,0)$.  We consider states with momentum and winding around the $\theta_5$ circle, which is unaffected by the orbifold.  In particular, when $p_{R,5}^2 \ge p_{L,5}^2 + P^2$ there are BPS states with $m^2 = \frac{2}{\alpha'} p_{R,5}^2$ on the sublattice $\Gamma_1 = \{P \in \Gamma_{16}| P\cdot e_i \in \mathbb{Z}\}$.

In fact, for generic $R_i$ these are on the only BPS states in the untwisted sector. For instance, states with $P\cdot e_1 \in \mathbb{Z}+1/2$ must have non-vanishing momentum on the $\theta_1$ circle, whereas states with $P\in\Gamma_{16} + e_1$ must have non-vanishing winding on the $\theta_1$ circle. In either case, the extra mass contribution from $p_{R,1}^2$ prevents the BPS bound $m^2 \ge \frac{2}{\alpha'} p_{R,5}^2$ from being saturated. For generic $R_i$, $\sum_{i=1}^4 p_{R,i}^2$ can only vanish in the absence of momentum and winding on $T^4$, hence none of these states are BPS.

The sublattice $\Gamma_1$ has coarseness $2$ and index $2^8 = 256$ inside the untwisted sector charge lattice $\Gamma_{\rm untw} = \Gamma_{16} + \Span(e_1,\ldots,e_4)$, which places a lower bound on the index and coarseness of the sublattice $\Gamma_{\rm ext}$ satisfying the sLWGC. In fact, there are superextremal states everywhere in the twisted sector coset $\Gamma_{\rm tw} = \Gamma_{\rm untw} + e_0$ because the problematic contribution $p_{R,i}^2$, $i=1,\ldots,4$, is absent. Thus, we can choose $\Gamma_{\rm ext} = \Gamma_1 \cup (\Gamma_1 + v)$ for any $v \in \Gamma_{tw}$, for a total of $256$ possible choices of $\Gamma_{\rm ext} \subset \Gamma = \Gamma_{\rm untw} \cup \Gamma_{\rm tw}$ satisfying the sLWGC, each with coarseness $2$ and index $256$.

One might expect that $T^4/\mathbb{Z}_N$ heterotic orbifolds with $N > 3$ would violate the LWGC even more badly, but this turns out to not be the case.  The reason is that Wilson lines for $N=4$ are constrained by the orbifold projection to lie in $\Gamma_{16} / 2$, so the sLWGC lattice in these examples has coarseness 2 rather than 4.  The index in this case is no larger than $2^2=4$, since Wilson lines on two of the $S^1$s are related in pairs.  For the case of $T^4/\mathbb{Z}_6$ orbifolds, Wilson lines are forbidden due to the orbifold action.

The coarseness of $\Gamma_{\rm ext}$ is (roughly speaking) a measure of the distance between lattice sites with superextremal particles. In the examples we have considered, this distance is never very large. On the other hand, the index of $\Gamma_{\rm ext}$ is (again, roughly speaking) a measure of the \emph{density} of lattice sites with superextremal particles. We have seen that this index can be large in specific examples. This suggests that the density of sites with superextremal particles can be rather low. However, this ignores the contribution of superextremal particles not on $\Gamma_{\rm ext}$. For instance, in the large index example above the fraction of lattice sites with a superextremal particle is actually more than half, including every twisted sector site. In cases such as these a different measure for how badly the LWGC is violated could be more physically relevant. We leave further consideration of these issues to future work.

\section{The sLWGC in \alt{AdS$_5$}{AdS5}: First Example}

We have argued that the sLWGC holds for closed string $U(1)$ gauge groups in perturbative string theory. It would be useful to have other examples, to help assess whether the sLWGC is likely to be a general property of quantum gravity theories or whether it is an artifact of the special cases studied so far. A rich landscape of examples to explore arises from the AdS/CFT correspondence \cite{nakayama:2015hga,Benjamin:2016fhe,Montero:2016tif}. For the moment we will take our working definition of a superextremal operator ${\cal O}$ in AdS$_5$ to be one for which 
\begin{equation}
  \frac{Q (\mathcal{O})}{\sqrt{b}} \geqslant \frac{\Delta
  (\mathcal{O})}{\sqrt{12 c}} \label{bound0}.
\end{equation}
This is a very direct translation of the flat-space extremality bound to the AdS context. Here $Q$ and $\Delta$ are the charge and dimension of the
operator in question, $c \sim \langle T T \rangle$ is the central charge of
the CFT, and $b \sim \langle J J \rangle$ is the beta function of the
conserved current. To fix the normalization, we note that for a free theory
\begin{align}
  b &= \sum_{\phi} k (\phi) Q (\phi)^2, \\
  c &= \frac{1}{120}  (12 n_v + 3 n_f + 2 n_s),
\end{align}
where $Q (\phi)$ is the charge of field $\phi$; $k
(\phi) = \frac{2}{3}$ for a Weyl fermion and $k (\phi) = \frac{1}{3}$ for a
complex scalar; and $n_s$, $n_f$ and $n_v$ are the number of complex scalars, Weyl fermions, and vectors in the theory. 

For theories with multiple conserved currents, the AdS$_5$ analog of the Convex Hull Condition is that the
charge-to-dimension vectors
\begin{equation}
  \zeta_a (\mathcal{O}) \equiv \frac{\sqrt{12 c}}{\Delta (\mathcal{O})} Q_a
  (\mathcal{O}) 
\end{equation}
satisfy the convex hull condition with the charge-space metric $b_{a b} \sim
\langle J_a J_b \rangle$, where
\begin{equation}
  b_{a b} = \sum_{\phi} k Q_a Q_b
\end{equation}
in a free theory. In practice we will pick a basis where $b_{a b}$ is
diagonal, so that $b_{a b} = \delta_{a b}$ after rescaling. When this is done,
we denote the rescaled $\vec{\zeta}$ as $\vec{Z}$ (implying a unit metric).

There are a number of subtleties in the translation of the Weak Gravity Conjecture to AdS \cite{nakayama:2015hga}: for instance, large AdS black holes have a very different extremality bound from flat-space black holes, and the translation of mass to $\Delta$ elides possible effects of the finite AdS curvature scale. For now, we will simply show that the naive bound (\ref{bound0}) is actually satisfied in full Lattice WGC form for the special case of AdS$_5 \times$ S$^5$ (up to a caveat about a small number of lattice sites associated with a free field). 

\subsection{\alt{$\mathcal{N}= 4$}{N=4} theories}

Consider an $\mathcal{N}= 4$ gauge theory with gauge
group $G$.\footnote{For simplicity, we assume that $G$ is a classical Lie
group.} In this case, all the global symmetries are contained within the $SU
(4)_R$ symmetry. For each vector multiplet there are four Weyl fermions in a
fundamental of $SU (4)_R$ and six real scalars in an antisymmetric
tensor of $SU (4)$ (which is a real representation, equivalent to the
vector of $SO (6)$). The $SU(4)_R$ charge eigenstates can be expressed in terms of the Cartan subalgebra:
\begin{equation}
  \begin{array}{c|ccccccc}
    & \psi_1 & \psi_2 & \psi_3 & \psi_4 & \phi_{14} = \phi_{23}^{\ast} &
    \phi_{24} = \phi_{31}^{\ast} & \phi_{34} = \phi_{12}^{\ast}\\ \hline
    U (1)_1 & \frac{1}{2} & - \frac{1}{2} & - \frac{1}{2} & \frac{1}{2} & 1 &
    0 & 0\\
    U (1)_2 & - \frac{1}{2} & \frac{1}{2} & - \frac{1}{2} & \frac{1}{2} & 0 &
    1 & 0\\
    U (1)_3 & - \frac{1}{2} & - \frac{1}{2} & \frac{1}{2} & \frac{1}{2} & 0 &
    0 & 1
  \end{array}
\end{equation}
We have
\begin{equation}
  b_{a b} = (\dim G) \begin{pmatrix}  
    1 & 0 & 0\\
    0 & 1 & 0\\
    0 & 0 & 1 \end{pmatrix} 
\end{equation}
in this basis, where the result is independent of the gauge coupling, hence
the same as in the free theory: $b_{a b} = \frac{2}{3}  \sum_f Q_{f a} Q_{f b}
+ \frac{1}{3}  \sum_s Q_{s a} Q_{s b}$. Likewise, we have $c = \frac{1}{4}
\dim G$ for the same reason, so that
\begin{equation}
  Z_a ({\rm Tr} \phi_{14}^2) = \left( \sqrt{3}, 0, 0 \right) \;  \;, \;  \;
  Z_a ({\rm Tr} \phi_{24}^2) = \left( 0, \sqrt{3}, 0 \right) \;  \;, \;  \;
  Z_a ({\rm Tr} \phi_{34}^2) = \left( 0, 0, \sqrt{3} \right) \,.
\end{equation}
Interestingly, the convex hull condition is satisfied by these three low-lying operators alone, and saturated in the direction $(1, 1, 1)$. However, there are infinitely many other charged operators, to which we now turn our attention.

\subsection{The charge lattice}

It is useful to rephrase the discussion in $\mathcal{N}= 1$ language. We have
the chiral multiplets
\begin{equation}
  \Psi_i = (\phi_{i 4}, \psi_i) 
\end{equation}
with the charge table:
\begin{equation}
  \begin{array}{c|c|ccc}
    & G & U (1)_X & U (1)_Y & U (1)_R\\ \hline
    \Psi_1 & \mathbf{\rm Adj} & 1 & 1 & 2 / 3\\
    \Psi_2 & \mathbf{\rm Adj} & - 1 & 1 & 2 / 3\\
    \Psi_3 & \mathbf{\rm Adj} & 0 & - 2 & 2 / 3
  \end{array}
\end{equation}
and the superpotential:
\begin{equation}
  W = {\rm Tr} \Psi_1 \Psi_2 \Psi_3 - {\rm Tr} \Psi_2 \Psi_1 \Psi_3 \,,
\end{equation}
where $U (1)_X = U (1)_1 - U (1)_2$, $U (1)_Y = U (1)_1 + U (1)_2 - 2 U (1)_3$
and $U (1)_R = \frac{2}{3}  \sum_i U (1)_i$. Here we show only the Cartan of
the global symmetry $SU (3) \times U (1)_R \subset SU (4)_R$
that is manifest in the $\mathcal{N}= 1$ description. The chiral ring is built
from the operators:
\begin{equation}
  \mathcal{O}_{i_1 \ldots i_n} \equiv {\rm Tr} [\Psi_{i_1} \ldots \Psi_{i_n}]
  \qquad (n \geqslant 2) \,,
\end{equation}
where $\mathcal{O}_{i_1 \ldots i_n}$ is completely symmetric in its indices by
the F-term constraints and the case $n=1$ vanishes identically for any semisimple gauge group $G$. (Below we will sometimes abuse notation by conflating chiral operators and their lowest components.)

The $\mathcal{N}= 1$ subalgebra described above implies that gauge-invariant
operators of the form $\phi_{14}^{n_1} \phi_{24}^{n_2} \phi_{34}^{n_3}$
(regardless of index structure) have vanishing anomalous dimension. Notice
that these operators cover the lattice of integer $U (1)_i$ charges for $Q_i
\geqslant 0$---apart from the points $(1, 0, 0)$, $(0, 1, 0)$ and $(0, 0,
1)$, due to the tracelessness of the adjoint representation---and satisfy~(\ref{bound0}).\footnote{To see that~(\ref{bound0}) is satisfied, it is sufficient to consider the contribution of the $U(1)_R$ charge, which saturates the bound in the absence of other charges.} The same is true for the operators:
\begin{equation}
  \phi_{23}^{n_1} \phi_{31}^{n_2} \phi_{34}^{n_3} \qquad, \qquad
  \phi_{31}^{n_1} \phi_{12}^{n_2} \phi_{14}^{n_3} \qquad, \qquad
  \phi_{23}^{n_1} \phi_{24}^{n_2} \phi_{12}^{n_3} \,,
\end{equation}
(and their conjugates), as these operators are chiral (anti-chiral) with
respect to other $\mathcal{N}= 1$ subalgebras. Together, these operators cover
the entire lattice of integer $U (1)_i$ charges---apart from $(\pm 1, 0, 0)$,
$(0, \pm 1, 0)$ and $(0, 0, \pm 1)$---and all satisfy (\ref{bound0}).

To get half-integer $U (1)_i$ charges, we must consider fermionic operators.
For instance, consider the chiral operator:
\begin{equation}
  \mathcal{O}^{\alpha}_{i_1 \ldots i_n} = \Tr [W^{\alpha} \Psi_{i_1} \ldots \Psi_{i_n}] =
  {\rm Tr} [\psi_4^{\alpha} \phi_{14}^{n_1} \phi_{24}^{n_2} \phi_{34}^{n_3}]\,,
  \qquad (n \geqslant 2)\,.
\end{equation}
This can easily be seen to satisfy~(\ref{bound0}), since $\psi_4$
lies in the $\langle \phi_{14}, \phi_{24}, \phi_{34} \rangle$ plane in
$Z$-space. Together with analogous operators for the other $\mathcal{N}= 1$
subalgebras and their complex conjugates, this covers all the half-integer
points on the charge lattice except $\left( \pm \frac{1}{2}, \pm \frac{1}{2},
\pm \frac{1}{2} \right)$.

To see that there are no gauge-invariant operators with the missing charges
$(\pm 1, 0, 0)$, $(0, \pm 1, 0)$, $(0, 0, \pm 1)$, or $\left( \pm \frac{1}{2},
\pm \frac{1}{2}, \pm \frac{1}{2} \right)$ which also satisfy~(\ref{bound0}), we
see that the bound requires $\Delta \leqslant \sqrt{3}$ for the
first three cases and $\Delta \leqslant \frac{3}{2}$ for the fourth. Since
bosonic (fermionic) operators have integer (half-integer) charges, the only
possible operators that can satisfy these constraints consistent with
unitarity bounds are scalars for the first three cases and $j = \left(
\frac{1}{2}, 0 \right)$ or $j = \left( 0, \frac{1}{2} \right)$ fermions for
the last. In the $g^2_{\rm YM} N \rightarrow 0$ limit, these operators must
have dimensions $\Delta = 1$ and $\Delta = \frac{3}{2}$, respectively,
indicating that they are built from a single fundmental field, but no such
gauge-invariant exists due to the tracelessness of the adjoint representation.

Thus, single-trace operators satisfying the WGC-like bound~(\ref{bound0}) cover nearly the entire charge
lattice. The missing sites are, curiously, precisely the components of a free
$\mathcal{N}= 4$ multiplet. In the case where $G=SU(N)$, these sites are restored by replacing $G \to U(N)$; the corresponding modes in $AdS_5$ are pure gauge within the bulk, hence the difference between the two cases comes down to boundary conditions~\cite{Aharony:1999ti}. Regardless of which choice we make, the sLWGC is satisfied with coarseness at most 2.

The cases $G=SO(N)$ or $G=U\!Sp(N)$, dual to orientifolds of $AdS_5 \times S^5$, are somewhat different. The pure gauge mode in the bulk is removed by the orientifold projection. Instead, there is a further constraint on the chiral ring coming from the constraint
\be
\Psi^\T = -M \Psi M^{-1} 
\ee
on the adjoint representation, where $M$ is the identity for $SO(N)$ and the symplectic form for $U\!Sp(N)$. Thus,
\be
\Tr [\Psi_{i_1} \ldots \Psi_{i_n}] = (-1)^n \Tr [M \Psi_{i_n} M^{-1} M \Psi_{i_{n-1}} M^{-1}\ldots M \Psi_{i_1} M^{-1}] = (-1)^n {\rm Tr} [\Psi_{i_1} \ldots \Psi_{i_n}] \,,
\ee
where we use the cyclic property of the trace and the F-term conditions in the second step. Thus, $\mathcal{O}_{i_1 \ldots i_n}$ vanishes identically for $n$ odd. By similar reasoning $\mathcal{O}^\alpha_{i_1 \ldots i_n}$ vanishes identically for $n$ even. The sLWGC is satisfied with coarseness of 2, as follows: consider any charge $Q$ in the $SU(4)_R$ charge lattice, where we can fix $Q$ to have non-negative entries without loss of generality after an $SU(4)_R$ transformation. If $Q$ is bosonic (has integer entries) then there is an operator of the form $\mathcal{O}_{i_1 \ldots i_{2n}}$ with charge $2Q$. If $Q$ is fermionic, then $Q$ can be written in the form $(\frac{1}{2},\frac{1}{2},\frac{1}{2})+P$ for $P$ bosonic with non-negative entries. Therefore, there is a superextremal operator of charge $2Q$ of the form
\be
\widetilde{\mathcal{O}}_{i_1 \ldots i_{2n}} \equiv \Tr( W^2 \Psi_{i_1} \ldots \Psi_{i_{2n}}) \,.
\ee
In the case $n=0$ this reduces to the exactly marginal operator which controls the gauge coupling.

\medskip

There is a large landscape of well-understood supersymmetric CFTs in which the validity of the Weak Gravity Conjecture, in both weak and strong forms, should be further explored. For now, we take this simplest example as encouraging additional evidence that a very strong form of the WGC is valid across a wide range of theories of quantum gravity.

\section{Further Remarks on the sLWGC}

\subsection{The sLWGC and UV cutoffs} \label{subsec:UVcutoffs}

Any variation of the Weak Gravity Conjecture that requires the existence of an infinite family of electrically charged particles automatically implies a cutoff on the validity of effective field theory. In contrast to the indirect argument of \cite{ArkaniHamed:2006dz} based on magnetic monopoles, the UV cutoff is immediately clear from the electric WGC in its lattice form: as we send the gauge coupling $e \to 0$, an entire tower of charged particles descends toward zero mass and any field theory based on a finite number of fields breaks down. Because a charge $q$ particle in this tower is constrained to have mass below $q e M_{\rm Pl}$, the bound on the validity of EFT is the same as that obtained from the magnetic argument: $\Lambda \simlt e k M_{\rm Pl}$ if $k$ is the coarseness of the sublattice satisfying the WGC.\footnote{Because at this time we have no examples where the coarseness $k$ is parametrically large, we will drop this factor in the remainder of the discussion. Large coarseness constitutes an interesting potential loophole that deserves further exploration.} This formulation of the UV cutoff bears a strong similarity to the original Swampland conjectures regarding the breakdown of EFT at long distances in moduli space \cite{Vafa:2005ui,Ooguri:2006in}.

A cutoff arising from a tower of fields does not necessarily imply an end to weakly-coupled physics: for instance, at energies above the scale of a Kaluza-Klein tower, we could switch to a local, weakly-coupled EFT description in higher dimensions. This is familiar, for instance, from the scenario of Large Extra Dimensions \cite{ArkaniHamed:1998rs, ArkaniHamed:1998nn}: in a sense, these theories break down at relatively long distances (microns to millimeters) where a tower of Kaluza-Klein gravitons is found. However, this is a fairly innocuous breakdown of effective field theory, since we know that experimentally such extra dimensions were perfectly consistent with data even after Standard Model scattering at the weak scale had been probed directly. The tower of light, weakly coupled modes has little effect on the dynamics of more strongly coupled fields until scattering processes reach sufficiently large energies. This larger energy scale is the higher-dimensional Planck scale, which determines the energy at which generic scattering processes are expected to form (higher-dimensional) black holes and short distance physics gives way to Asymptotic Darkness \cite{Banks:1999gd,Giddings:2001bu,Banks:2003vp}.

The Lattice WGC implies that {\em any} weakly-coupled gauge theory behaves roughly like an extra dimension, in the sense that it lowers the fundamental scale $\Lambda_*$ at which gravity becomes strong (and all scattering processes become strongly coupled).\footnote{One definition of $\Lambda_*$ is the smallest radius for which we can talk about a black hole. It is not necessarily the center-of-mass energy at which generic scattering processes produce black holes and Asymptotic Darkness sets in, since semiclassical $D$-dimensional black holes have energy at least $M_D^{D-2}/\Lambda_*^{D-3}$. (However, there may more generally be states analogous to higher-dimensional black holes near the scale $\Lambda_*$.) MR thanks Emil Martinec for pointing out that the phrasing in the first preprint version of this paper was flawed.} The reason is the familiar species bound \cite{ArkaniHamed:2005yv,Distler:2005hi,Dimopoulos:2005ac,Dvali:2007hz,Dvali:2007wp}, which states that the strong gravity scale is parametrically below the Planck scale in theories with a large number of species $N_{\rm dof}(\Lambda_*)$ that are light compared to $\Lambda_*$. In a theory of $D$ large dimensions,
\be
\Lambda_*^{D-2} \simlt \frac{M_{\rm Pl}^{D-2}}{N_{\rm dof}(\Lambda_*)}.   \label{eq:species}
\ee
If the degrees of freedom that we are counting correspond to Kaluza-Klein modes, this gives the familiar matching between the lower-dimensional Planck scale $M_{\rm Pl}$ and the higher-dimensional one $\Lambda_*$. But it is more general. One appealing argument for (\ref{eq:species}) is that a black hole of radius $\Lambda_*^{-1}$ will evaporate extremely rapidly by Hawking radiation due to the many species to which it can decay, and vanish in a time shorter than its light-crossing time if the bound is violated (see \S3.1 of \cite{Dvali:2007wp}). This shows that the bound is necessary in order for the whole notion of semiclassical GR at the distance scale $\Lambda_*^{-1}$ to be coherent.

Let us combine the species bound (\ref{eq:species}) with the Lattice WGC, which tells us that in a U(1) gauge theory with coupling constant $e$ we have a number of charged particles
\be
N_{\rm dof}(\Lambda_*) \simgt \frac{\Lambda_*}{e M_{\rm Pl}^{(D-2)/2}}.
\ee
Combining the two inequalities we learn that
\be
\Lambda_* \simlt e^{1/(D-1)} M_{\rm Pl}^{3(D-2)/(2(D-1))}.
\ee
In the four-dimensional case this is simply $\Lambda_* \simlt e^{1/3} M_{\rm Pl}$. Such a relationship is in principle empirically falsifiable. Because we know that strong gravitational scattering has not yet set in at LHC energies of around 1 TeV, the Lattice WGC together with the species bound imply that we will never observe a gauge theory with coupling $e \simlt (1~{\rm TeV}/M_{\rm Pl})^3 \sim 10^{-45}$. Testing such tiny couplings is not possible in the foreseeable future, but we can draw other conclusions. For instance, equivalence principle tests tell us that any massless $B-L$ gauge boson must have a coupling $e_{B-L} \simlt 10^{-24}$ \cite{Wagner:2012ui,Heeck:2014zfa}. From this we conclude that {\em if} a future experiment detected a long range $B-L$ force, the fundamental cutoff scale of gravity in our universe would necessarily be $\Lambda_* \simlt 10^{10}~{\rm GeV}$. Such a low fundamental cutoff would have many implications, for instance for models of unification, neutrino masses, supersymmetry breaking, and inflation. In particular, the energy density $V_{\rm inf}$ during inflation would be below this scale, implying an infinitesimal tensor-to-scalar ratio $r \propto V_{\rm inf}$ that would never be measured. Thus the discovery of {\em both} a $B-L$ gauge boson in fifth force searches {\em and} primordial tensor modes in the CMB would falsify the Lattice WGC. (We have set aside for the moment the question of whether the WGC forbids super-Planckian field ranges and hence detectable tensors; inflation with a field range {\em near} $M_{\rm Pl}$, rather than much larger, is likely compatible with the WGC with modest tuning in any case.)

Just as with extra dimensions, if there are {\em multiple} weakly-coupled gauge groups with small couplings, the constraints on the scale of strong gravity can become stronger. For nonabelian groups it is the Cartan generators that matter for counting charges in lattice versions of the WGC, so SU(2) will lead to the same bound as a single U(1) but larger SU($N$) groups are more highly constrained.\footnote{This is a conservative upper bound on the cutoff which we believe can be improved. In the present analysis, we have not accounted for the fact that the charged states must come in complete representations of the non-Abelian gauge group $G$ with equal masses for each component. The consequences of this requirement depend somewhat on further assumptions about the spectrum. If we impose the slightly stronger requirement that there is a superextremal particle in every representation of $G$ (or an appropriate sub-lattice variant of this) then we find, e.g., $\Lambda_* \lesssim g^{1/2} M_{\rm Pl}$ for $G=\text{SU}(2)$. We discuss these points in more detail in a forthcoming paper.}
 For an SU($N$) gauge group with coupling $g$ in four dimensions, the bound is 
\be
\Lambda_* \simlt g^{(N-1)/(N+1)} M_{\rm Pl}. \label{eq:nonabelianLambdastar}
\ee
In the large $N$ limit strong gravity sets in immediately at the scale where the tower of charged modes begins. For moderate values of $N$ there is an interesting potential point of contact with current and near-future cosmological observations. Dark matter coupled to nonabelian dark radiation can damp the matter power spectrum in a specific way \cite{Buen-Abad:2015ova,Lesgourgues:2015wza,Cyr-Racine:2015ihg} that has been argued to improve fits to observation by $3\sigma$ \cite{Lesgourgues:2015wza}. The detectable effect occurs for small couplings $g \sim 2 \times 10^{-4}$. For the case of $N \simgt 4$, such a small coupling in (\ref{eq:nonabelianLambdastar}) leads to nontrivial bounds on the tensor-to-scalar ratio $r$, e.g.~$r \simlt 0.01$ for $N \geq 5$. There is no obvious direct test of the value of $N$, and $N = 2$ or $N = 3$ would be compatible with both the existence of weakly-coupled dark radiation at a detectable level and a sizable tensor-to-scalar ratio, so this does not quite constitute a sharp empirical test of the Lattice WGC. But it is tantalizingly close to one, and suggests that further scrutiny of the Lattice WGC may lead to interesting phenomenological tests.

\subsection{A comment on Higgsing}\label{subsec:higgsing}

It has been pointed out that the LWGC is not robust under Higgsing the gauge group, which produces smaller couplings in the surviving infrared gauge group than were present in the ultraviolet \cite{Saraswat:2016eaz}. It turns out that the sLWGC is also not robust under Higgsing. This suggests two possible viewpoints. One is that the WGC (in any of its forms) is a statement about ultraviolet physics that requires knowledge of the full un-Higgsed charge lattice, and cannot constrain effective field theories \cite{Saraswat:2016eaz}. Such a view, if correct, invalidates the argument we have made about a fundamental cutoff based on measured gauge couplings. The second possibility is that the WGC applies to a theory in any of its infrared phases, in which case we should seek a still more robust statement about UV physics to guarantee this. In this case, our arguments would apply. One point in favor of this latter viewpoint is that the distinction between a Higgsed gauge group and a massive stringy U(1) is not completely sharp, so the meaning of the ``full charge lattice'' is ambiguous, whereas the gauge group in the infrared is well-defined; this is simply a reflection of the widely accepted belief that quantum gravities are not field theories, hence there is no UV notion of the gauge group.

This distinction could be very important: for example, the extra-dimensional version of the ``clockwork axion'' model \cite{Kaplan:2015fuy} involves a sequence of $N$ Higgs fields used to produce an effective gauge coupling that is exponentially suppressed (and, ultimately, field range that is exponentially large) in $N$.\footnote{The idea of realizing a $\bb{Z}/(3^N)$ symmetry with $N$ scalar fields coupled as $\phi_i^\dagger \phi_{i+1}^3$ appeared already in \cite{Dvali:2007hz}.} If the sLWGC indeed constrains IR physics, any such theory necessarily comes with an exponentially low cutoff scale; if the sLWGC constrains only the UV theory before Higgsing, the constraint is very mild. The question this raises is whether, for a polynomial cost in the UV cutoff or the number of light fields, one can obtain gauge couplings that are exponentially small or merely polynomially small. In the latter (``UV-only'') view of the WGC, quantum gravity puts up very little resistance to attempts to produce approximate continuous global symmetries.

If the viewpoint of~\cite{Saraswat:2016eaz} is correct, then there should be examples of quantum gravities which satisfy the WGC in some UV description but which fail to satisfy it after Higgsing. What we can say for now is that the modular invariance argument applies equally well on Higgs branches, depending only on the conserved currents of the worldsheet theory (which are the {\em surviving} massless gauge bosons in spacetime). As a cross-check, we have found by explicit computation that the sLWGC is satisfied for small VEVs on the Higgs branch of a $T^4/\bb{Z}_3$ orbifold of the heterotic string with SU(9) gauge group. Our expectation is that consistent quantum gravities always conspire to satisfy the WGC in the infrared, but fully addressing
 this UV-versus-IR interpretational question about the nature of the WGC is probably the most urgent task ahead for phenomenological applications.

\subsection{The sLWGC and stable particles}

Thus far, we have allowed unstable resonances to satisfy the sLWGC.  This presents an ambiguity in the definition of some forms of the WGC, since the mass of a resonance, specified by the location of a pole in the complex-energy plane of the S-matrix, is a complex quantity.  For a weakly coupled theory, where the decay width is small, this ambiguity is relatively benign.  For a strongly coupled theory, it is a serious issue and requires a revised statement of the sLWGC in terms of stable states.

As stated, the sLWGC implies a sharp statement regarding stable states: for any spot $\vec{q}$ in the sublattice $\Gamma_{\rm ext} \subseteq \Gamma$ of states satisfying the sLWGC, there must exist a (possibly multiparticle) superextremal state.  Once again, note that we are referring to the mass of a multiparticle state as the sum of the masses of the individual particles, which is a valid measure of the energy carried away by the state as long as interactions between the particles die off at infinity.  If one wishes to exclude resonances from the discussion entirely, the existence of a finite-coarseness sublattice $\Gamma_{\rm ext} \subseteq \Gamma$ of superextremal (possibly multiparticle) states can even be taken as the definition of the sLWGC.\footnote{This reframing is not necessary in the special case where the extremality bound is a BPS bound, as there cannot be any superextremal resonances.}

Resonances necessarily play an important role in the sLWGC, however, since KK modes of superextremal multiparticle states are generically not superextremal, whereas KK modes of superextremal resonances are \cite{Heidenreich:2015nta}.  Indeed, KK modes of unstable resonances can even be stable particles.  For the sLWGC to be consistent under KK reduction on a circle, resonances are generically required.

\subsection{Possible sLWGC variants}

It is interesting to speculate about possible variants of the Weak Gravity Conjecture in addition to the sLWGC.  In the examples discussed here that violate the LWGC, we note that the charge sites without a superextremal particle are rather sparse.  Indeed, these sites always lie in a proper sublattice of the charge lattice.  This implies that the charges of superextremal particles generate the entire charge lattice.  This latter statement is known to be true for M-theory compactifications on smooth Calabi-Yau three-folds: the integral Hodge conjecture, proven by Voisin for Calabi-Yau three-folds \cite{voisin2004integral}, shows that holomorphic curve classes generate the full homology lattice $H_2$.  Equivalently, BPS states of M2-branes wrapped on holomorphic curves generate the entire electric charge lattice.

Several toroidal orbifold constructions na\"ively violate these conjectures, but they involve subtleties making them unconvincing counterexamples.  Consider first type II string theory on $T^7 / \bb{Z}_2 \times \bb{Z}_2$ with generators\footnote{We thank Cumrun Vafa for pointing out this example to us.}
\begin{equation}
\theta: e_1 \rightarrow -e_1 + \frac{1}{2},\, e_{2,3,4} \rightarrow - e_{2,3,4}\,, \qquad \omega: e_{1,2,5,6} \rightarrow - e_{1,2,5,6},\, e_{3,7} \rightarrow  e_{3,7} + \frac{1}{2} \,,
\end{equation}
which defines a gravitational theory in three dimensional Minkowski space. Strictly speaking, black holes do not exist in this theory, in part because adding energy to the vacuum leads to a deficit angle at infinity: since the deficit angle cannot exceed $2 \pi$, masses cannot be parametrically super-Planckian. Nonetheless, we naively apply the Weak Gravity Conjecture if we normalize the extremality bound using the large charge asymptotics of the tree-level string theory spectrum.
Doing so, we find a violation to the aforementioned WGC variants.  In the $\omega$-twisted sector, there are no fixed points, and all states have half-integral winding number around the $e_3$ direction.  In the $\omega \theta$-twisted sector, all states have half-integral winding number around the $e_1$ direction.  Thus, all states of half-integral winding charge around $e_7$ are subextremal, and superextremal states do not generate the full charge lattice. 
 By considering M-theory on the same orbifold, we find the same results for a 4d theory with eight supercharges involving strings rather than particles. However, strings in four dimensions generate a deficit angle, much like particles in the three dimensions, and neither ``counterexample'' is convincing or even necessarily addressing a well formulated conjecture.

For a possible counterexample to these conjectures involving particles in four dimensions, consider heterotic string theory on the same $T^6/ \bb{Z}_2 \times \bb{Z}_2$ orbifold,
\begin{equation}
\theta: e_1 \rightarrow -e_1 + \frac{1}{2},\, e_{2,3,4} \rightarrow - e_{2,3,4}\,, \qquad \omega: e_{1,2,5,6} \rightarrow - e_{1,2,5,6},\, e_{3} \rightarrow  e_{3} + \frac{1}{2}\,.
\end{equation}
The set of superextremal states in this orbifold construction depends on the radii of the $T^6$.  In the $\omega$-twisted sector, there are no fixed points, and all states have half-integral winding number around the $e_3$ direction.  In the $\omega \theta$-twisted sector, all states have half-integral winding around the $e_1$ direction.  If the radii of these two circles are fixed at $R_1^2 , R_3^2 > 12 \alpha'$ then all the states in these subsectors are subextremal, and superextremal states will not generate the charge lattice.  However, since this theory has only four real supercharges, the radions will generically be stabilized by quantum corrections.  It is plausible that vacua with $R_1^2, R_3^2$ greater than $12 \alpha'$ do not exist---or that quantum corrections significantly modify the spectrum---so this is not a clear counterexample to these variants of the conjecture either.

It would be desirable to find a more convincing counterexample to these conjectures, or else to put them on more solid footing.

\section{Conclusions}

In previous work \cite{Heidenreich:2015nta} we pointed out that the original Weak Gravity Conjecture is not robust under compactification: a theory that satisfies the conjecture can give rise in the infrared to other theories that fail it. We proposed that a stronger conjecture could be more robust, suggesting the Lattice Weak Gravity Conjecture as a natural generalization that passed some preliminary checks. Here we have seen that although our original statement of the Lattice Weak Gravity Conjecture was too strong, the Sublattice Weak Gravity Conjecture passes many tests and follows from modular invariance in perturbative string theory (assuming that the asymptotics of the perturbative string spectrum and of semiclassical black holes match, which is known in many cases but we have not proven in general). This lattice form of the conjecture is also very closely related to Swampland conjectures regarding effective field theory on moduli spaces \cite{Vafa:2005ui,Ooguri:2006in}: when a gauge coupling is sent to zero, we approach a point at infinite distance on moduli space and a tower of particles becomes light, invalidating the EFT.

There are many concrete directions for further work that can shed light on the WGC and its applications. A major question raised by the Sublattice Weak Gravity Conjecture is how coarse the sublattice can be: can the original Lattice Weak Gravity Conjecture fail badly? If the coarseness of the sLWGC-satisfying sublattice in the full charge lattice is very large, there may be no light superextremal particles and applications to inflation will be limited. However, in the examples that we understand, the coarseness of the sublattice is small. These examples are limited by supersymmetry and perturbative control. Interesting possibilities are that either there is some fundamental upper bound on the index (or possibly the coarseness) of the sublattice or, perhaps more plausibly, that sparse sublattices might be achieved in some nonsupersymmetric examples but at the cost of lowering the cutoff scale of the theory. We have seen before that combinations of Weak Gravity arguments with arguments about low cutoffs on effective field theory have the potential to be quite powerful in constraining large-field inflation models \cite{Heidenreich:2015wga}, so this could be an interesting outcome. Note also that the quotient of the full charge lattice by the sLWGC sublattice is a discrete group. It is tempting to suspect that some other physics associated to this discrete group could help to explain why the lightest particles at these lattice sites are subextremal.

The sLWGC, like the original LWGC, also suggests that if we go far out on the charge lattice we should find superextremal black holes. This may at first sound like a contradiction in terms, but corrections to semiclassical GR from higher-dimension operators indeed can produce black holes that are superextremal as viewed relative to the asymptotic linear charge-to-mass relationship \cite{kats:2006xp}. If one could show that the corrections to the charge-to-mass relationship have a definite sign, it would both prove the most mild form of the WGC and offer further supporting evidence for the LWGC. We plan to present some results in this direction in a future publication.

The sLWGC, if true in general quantum gravity theories, would have important implications. We have previously sketched an argument that the LWGC would rule out super-Planckian field ranges in the simplest $N$-flation scenarios \cite{Heidenreich:2015wga}. It is likely that many other axion inflation theories are strongly constrained by the existence of an infinite tower of charged particles, contributing an infinite set of instanton terms whose size is bounded by the conjecture. Furthermore, we have argued that the sLWGC in general implies an ultraviolet cutoff on the validity of effective field theory---both a mild cutoff, associated with a tower of charged particles, and a more severe cutoff associated with strong gravity. This can be used to put phenomenological constraints on the existence of very weakly coupled gauge groups in our universe (up to the caveat discussed in \S\ref{subsec:higgsing}).

The connection between the sLWGC and modular invariance is intriguing. It suggests that further exploration of modular invariance in higher dimensions in the AdS/CFT context could lead to new insights, although the state/operator correspondence prefers sphere compactifications to torus compactifications and so some new ideas are likely needed to make progress. 

The accumulation of new examples, and the general proof for perturbative string theory that we have presented, offer further evidence for the validity of a strong version of the Weak Gravity Conjecture. Still, a more solid argument in favor of the conjecture on completely general grounds would be welcome. Some hopes for such arguments come from black hole physics and entropy bounds \cite{ArkaniHamed:2006dz,banks:2006mm}, factorization in spacetimes with multiple asymptotic boundaries \cite{harlow:2015lma}, AdS/CFT \cite{nakayama:2015hga,Benjamin:2016fhe,Montero:2016tif}, the Complexity Equals Action conjecture \cite{Brown:2015bva,Brown:2015lvg}, or Cosmic Censorship \cite{Horowitz:2016ezu}. We hope that the lively activity surrounding these questions will lead to new insights on quantum gravity, cosmology, and particle physics.

\section*{Acknowledgments}
We thank Alexander Maloney, Liam McAllister, Miguel Montero, David Morrison, Prashant Saraswat, Gary Shiu, Cumrun Vafa, Shing-Tung Yau, and Emil Martinec for discussions or correspondence. BH was supported in part by the Fundamental Laws Initiative of the Harvard Center for the Fundamental Laws of Nature and in part by Perimeter Institute for Theoretical Physics. Research at Perimeter Institute is supported by the Government of Canada through Industry Canada and by the Province of Ontario through the Ministry of Economic Development and Innovation. MR was supported in part by the NSF Grant PHY-1415548. TR was supported by NSF grant PHY-1067976 and by the NSF GRF under DGE-1144152. TR and BH would like to thank the 2016 Workshop on the Weak Gravity Conjecture and Cosmology and the Universidad Autonoma de Madrid for their hospitality during the early stages of this project.  MR would like to thank the String-Pheno 2016 conference and the University of Ioannina for their hospitality while this work was completed.

\appendix

\section{Modular Invariance and Chemical Potential for Rotations}
\label{app:modinvrot}

A weak point of our modular invariance argument for the sLWGC in \S\ref{subsec:modularinvariance} is that in a supersymmetric theory the modular invariant partition function $Z(\mu,\tau)$ has an insertion of $(-1)^F$ and is in fact zero. This is a technical subtlety that can be remedied by computing any other quantity that also has nice modular transformation properties, but which distinguishes between bosons and fermions and hence is nonzero. One way to do this is to observe that bosons and fermions transform differently under spacetime symmetries, so if we turn on an {\em additional} chemical potential for rotations in uncompactified directions in the target space we can obtain a nonvanishing quantity. Our goal in this appendix is to demonstrate that turning on such a chemical potential is possible for non-compact bosons, and gives rise to a modular covariant partition function. In particular, because the fields are noncompact there is a subtlety from zero modes that requires careful regulation.

Consider a theory with two free bosons, combined into a complex free boson $W
= X + i Y$. In addition to translations, we can now turn on a chemical
potential for rotations $W \rightarrow e^{i \theta} W$. From the oscillators, we get
\begin{equation}
  Z_{\rm osc} (\mu, \tau) = (q \bar{q})^{- \frac{1}{12}} \prod_{j =
  1}^{\infty} \frac{1}{1 - t q^j}  \frac{1}{1 - t^{- 1} q^j}  \frac{1}{1 -
  \bar{t}  \bar{q}^j}  \frac{1}{1 - \bar{t}^{- 1}  \bar{q}^j},
\end{equation}
where $t = e^{2 \pi i \mu}$, $\bar{t} = e^{- 2 \pi i \bar{\mu}}$. However, we
still need to account for the zero modes, which is tricky because the momentum
eigenstates transform non-trivially under rotations. In general, we can
Fourier transform
\begin{equation}
  | k, n \rangle = \int_0^{2 \pi} e^{i n \phi}  | k, \phi \rangle \dd \phi,
\end{equation}
where $| k, \phi \rangle$ denotes a momentum eigenstate with $\vec{k} = k
\hat{k}$ in the $\phi$ direction. Thus, the zero modes give us states of every
integer charge. However, we still need to account for the density of states. For a finite spatial volume $V$ we have $\dd \rho \sim V \frac{k \dd k}{2 \pi}.$
The more states at a given $k$, the higher the charge that can be formed by a
discrete Fourier transform. Thus, the charge is limited to  $| n
| \lesssim \sqrt{V} k / 2$, hence
\begin{equation}
  Z_0 \simeq \sum_n t^n \int_{2 | n | / \sqrt{V}}^{\infty} e^{- \pi \alpha'
  \tau_2 k^2}  \left( \sqrt{V}  \frac{\dd k}{2 \pi} \right) = \sqrt{V} 
  \sum_n \left( \frac{t^n}{4 \pi \sqrt{\alpha' \tau_2}} \right) {\rm erfc}
  \left( 2 | n |  \sqrt{\frac{\pi \alpha' \tau_2}{V}} \right) .
\end{equation}
Notice that if we set $t = 1$ and take the limit $V \gg \alpha' \tau_2$, we
recover the usual result $Z_0 \rightarrow V/(4 \pi^2 \tau_2)$.
If we instead hold $t$ fixed and take the limit $V \rightarrow \infty$, we get
\begin{equation}
  Z_0 \rightarrow \sum_n \left( \frac{\sqrt{V}}{4 \pi \sqrt{\tau_2}} - \frac{|
  n |}{\pi} \right) t^n .
\end{equation}
However, 
$\sum_n t^n = \sum_{m = - \infty}^{\infty} \delta (\mu - m),$
so for generic $\mu$ we can ignore the first term, which gives
\begin{equation}
  Z_0 = - \frac{1}{\pi}  \sum_{n = 1}^{\infty} n (t^n + t^{- n}) =
  \frac{2}{\pi} \cdot \frac{1}{(1 - t) (1 - t^{- 1})} = \frac{1}{2 \pi (\sin
  \pi \mu)^2} .
\end{equation}
So far, we have not distinguished between left and right and we have kept $\mu
= \bar{\mu}$. This generalizes to
\begin{equation}
  Z_0 = \frac{1}{2 \pi (\sin \pi \mu) (\sin \pi \bar{\mu})},
\end{equation}
which will turn out to be consistent with modular invariance.

Putting together the pieces, we obtain:
\begin{equation}
  Z (\mu, \tau) = \frac{2}{\pi} \cdot \frac{| \eta (\tau) |^2}{|
  \vartheta_{11} (\mu, \tau) |^2}, \label{appeq:Zmutau}
\end{equation}
where $\vartheta_{11} (\mu, \tau)$ is one of the Jacobi theta functions, with
the product representation
\begin{equation}
  \vartheta_{11} (\mu, \tau) = - 2 q^{1 / 8} \sin \pi \mu \prod_{m = 1}^{\infty} (1 - q^m) (1 - t q^m) (1 - t^{-1} q^m)\,.
\end{equation}
Under modular transformations $\vartheta_{11} (\mu, \tau + 1) = e^{\pi i / 4}
\vartheta_{11} (\mu, \tau)$ and
\begin{equation}
  \vartheta_{11} (\mu / \tau, - 1 / \tau) = - i (- i \tau)^{1 / 2} e^{\pi i
  \mu^2 / \tau} \vartheta_{11} (\mu, \tau) .
\end{equation}
Comparing with the modular transformation of $\eta (\tau)$, we see that $Z
(\mu, \tau)$ is modular covariant with $k = \tilde{k} = - 1$. This is a bit
pathological, because ordinarily $k, \tilde{k}$ would be positive by unitarity; it seems to be a consequence of infinite volume and the need for a regulator. Notice that $Z (\mu, \tau) \rightarrow \infty$ as $\mu \rightarrow 0$ or
$\bar{\mu} \rightarrow 0$. This is expected behavior, as $Z \propto V$ in the
case without a chemical potential and we need to take $V \rightarrow \infty$
to get a continuous rotational symmetry.

To summarize, turning on a chemical potential for spacetime rotations produces a partition function transforming in a simple modular covariant way. The subtleties introduced by noncompact fields are dealt with by regulating with finite volume. It can be checked by explicit calculation that the supersymmetric partition function in the presence of such chemical potentials does not vanish: as in (\ref{appeq:Zmutau}), the Jacobi theta functions appearing in the standard formulas become functions of the additional spacetime chemical potentials. If we turn on generic chemical potentials, the abstruse identity no longer enforces a cancelation. By turning on chemical potentials for both $U(1)$ symmetries and spacetime rotations, we can apply the modular invariance argument given in \S\ref{subsec:modularinvariance} to supersymmetric theories.

\bibliographystyle{jhep}
\bibliography{ref}

\providecommand{\href}[2]{#2}\begingroup\raggedright\begin{thebibliography}{10}

\bibitem{Vafa:2005ui}
C.~Vafa, \emph{{The String landscape and the swampland}},
  \href{http://arxiv.org/abs/hep-th/0509212}{{\tt hep-th/0509212}}.

\bibitem{Ooguri:2006in}
H.~Ooguri and C.~Vafa, \emph{{On the Geometry of the String Landscape and the
  Swampland}},
  \href{http://dx.doi.org/10.1016/j.nuclphysb.2006.10.033}{\emph{Nucl.Phys.}
  {\bf B766} (2007) 21--33}, [\href{http://arxiv.org/abs/hep-th/0605264}{{\tt
  hep-th/0605264}}].

\bibitem{ArkaniHamed:2006dz}
N.~Arkani-Hamed, L.~Motl, A.~Nicolis and C.~Vafa, \emph{{The String landscape,
  black holes and gravity as the weakest force}},
  \href{http://dx.doi.org/10.1088/1126-6708/2007/06/060}{\emph{JHEP} {\bf 0706}
  (2007) 060}, [\href{http://arxiv.org/abs/hep-th/0601001}{{\tt
  hep-th/0601001}}].

\bibitem{Bachlechner}
T.~C. Bachlechner, \emph{{Private communication}},  2015.

\bibitem{banks:2006mm}
T.~Banks, M.~Johnson and A.~Shomer, \emph{{A Note on Gauge Theories Coupled to
  Gravity}}, \href{http://dx.doi.org/10.1088/1126-6708/2006/09/049}{\emph{JHEP}
  {\bf 0609} (2006) 049}, [\href{http://arxiv.org/abs/hep-th/0606277}{{\tt
  hep-th/0606277}}].

\bibitem{cheung:2014ega}
C.~Cheung and G.~N. Remmen, \emph{{Infrared Consistency and the Weak Gravity
  Conjecture}}, \href{http://dx.doi.org/10.1007/JHEP12(2014)087}{\emph{JHEP}
  {\bf 1412} (2014) 087}, [\href{http://arxiv.org/abs/1407.7865}{{\tt
  1407.7865}}].

\bibitem{delaFuente:2014aca}
A.~de~la Fuente, P.~Saraswat and R.~Sundrum, \emph{{Natural Inflation and
  Quantum Gravity}},
  \href{http://dx.doi.org/10.1103/PhysRevLett.114.151303}{\emph{Phys.Rev.Lett.}
  {\bf 114} (2015) 151303}, [\href{http://arxiv.org/abs/1412.3457}{{\tt
  1412.3457}}].

\bibitem{nakayama:2015hga}
Y.~Nakayama and Y.~Nomura, \emph{{Weak gravity conjecture in the AdS/CFT
  correspondence}},
  \href{http://dx.doi.org/10.1103/PhysRevD.92.126006}{\emph{Phys. Rev.} {\bf
  D92} (2015) 126006}, [\href{http://arxiv.org/abs/1509.01647}{{\tt
  1509.01647}}].

\bibitem{harlow:2015lma}
D.~Harlow, \emph{{Wormholes, Emergent Gauge Fields, and the Weak Gravity
  Conjecture}}, \href{http://dx.doi.org/10.1007/JHEP01(2016)122}{\emph{JHEP}
  {\bf 01} (2016) 122}, [\href{http://arxiv.org/abs/1510.07911}{{\tt
  1510.07911}}].

\bibitem{Horowitz:2016ezu}
G.~T. Horowitz, J.~E. Santos and B.~Way, \emph{{Evidence for an Electrifying
  Violation of Cosmic Censorship}},
  \href{http://dx.doi.org/10.1088/0264-9381/33/19/195007}{\emph{Class. Quant.
  Grav.} {\bf 33} (2016) 195007}, [\href{http://arxiv.org/abs/1604.06465}{{\tt
  1604.06465}}].

\bibitem{Heidenreich:2015nta}
B.~Heidenreich, M.~Reece and T.~Rudelius, \emph{{Sharpening the Weak Gravity
  Conjecture with Dimensional Reduction}},
  \href{http://dx.doi.org/10.1007/JHEP02(2016)140}{\emph{JHEP} {\bf 02} (2016)
  140}, [\href{http://arxiv.org/abs/1509.06374}{{\tt 1509.06374}}].

\bibitem{Rudelius:2014wla}
T.~Rudelius, \emph{{On the Possibility of Large Axion Moduli Spaces}},
  \href{http://dx.doi.org/10.1088/1475-7516/2015/04/049}{\emph{JCAP} {\bf 1504}
  (2015) 049}, [\href{http://arxiv.org/abs/1409.5793}{{\tt 1409.5793}}].

\bibitem{Bachlechner:2014gfa}
T.~C. Bachlechner, C.~Long and L.~McAllister, \emph{{Planckian Axions in String
  Theory}}, \href{http://dx.doi.org/10.1007/JHEP12(2015)042}{\emph{JHEP} {\bf
  12} (2015) 042}, [\href{http://arxiv.org/abs/1412.1093}{{\tt 1412.1093}}].

\bibitem{Rudelius:2015xta}
T.~Rudelius, \emph{{Constraints on Axion Inflation from the Weak Gravity
  Conjecture}},
  \href{http://dx.doi.org/10.1088/1475-7516/2015/9/020}{\emph{JCAP} {\bf 09}
  (2015) 020}, [\href{http://arxiv.org/abs/1503.00795}{{\tt 1503.00795}}].

\bibitem{Montero:2015ofa}
M.~Montero, A.~M. Uranga and I.~Valenzuela, \emph{{Transplanckian axions!?}},
  \href{http://dx.doi.org/10.1007/JHEP08(2015)032}{\emph{JHEP} {\bf 08} (2015)
  032}, [\href{http://arxiv.org/abs/1503.03886}{{\tt 1503.03886}}].

\bibitem{Brown:2015iha}
J.~Brown, W.~Cottrell, G.~Shiu and P.~Soler, \emph{{Fencing in the Swampland:
  Quantum Gravity Constraints on Large Field Inflation}},
  \href{http://dx.doi.org/10.1007/JHEP10(2015)023}{\emph{JHEP} {\bf 10} (2015)
  023}, [\href{http://arxiv.org/abs/1503.04783}{{\tt 1503.04783}}].

\bibitem{Bachlechner:2015qja}
T.~C. Bachlechner, C.~Long and L.~McAllister, \emph{{Planckian Axions and the
  Weak Gravity Conjecture}},
  \href{http://dx.doi.org/10.1007/JHEP01(2016)091}{\emph{JHEP} {\bf 01} (2016)
  091}, [\href{http://arxiv.org/abs/1503.07853}{{\tt 1503.07853}}].

\bibitem{Hebecker:2015rya}
A.~Hebecker, P.~Mangat, F.~Rompineve and L.~T. Witkowski, \emph{{Winding out of
  the Swamp: Evading the Weak Gravity Conjecture with F-term Winding
  Inflation?}},
  \href{http://dx.doi.org/10.1016/j.physletb.2015.07.026}{\emph{Phys. Lett.}
  {\bf B748} (2015) 455--462}, [\href{http://arxiv.org/abs/1503.07912}{{\tt
  1503.07912}}].

\bibitem{Brown:2015lia}
J.~Brown, W.~Cottrell, G.~Shiu and P.~Soler, \emph{{On Axionic Field Ranges,
  Loopholes and the Weak Gravity Conjecture}},
  \href{http://dx.doi.org/10.1007/JHEP04(2016)017}{\emph{JHEP} {\bf 04} (2016)
  017}, [\href{http://arxiv.org/abs/1504.00659}{{\tt 1504.00659}}].

\bibitem{Junghans:2015hba}
D.~Junghans, \emph{{Large-Field Inflation with Multiple Axions and the Weak
  Gravity Conjecture}},
  \href{http://dx.doi.org/10.1007/JHEP02(2016)128}{\emph{JHEP} {\bf 02} (2016)
  128}, [\href{http://arxiv.org/abs/1504.03566}{{\tt 1504.03566}}].

\bibitem{Heidenreich:2015wga}
B.~Heidenreich, M.~Reece and T.~Rudelius, \emph{{Weak Gravity Strongly
  Constrains Large-Field Axion Inflation}},
  \href{http://dx.doi.org/10.1007/JHEP12(2015)108}{\emph{JHEP} {\bf 12} (2015)
  108}, [\href{http://arxiv.org/abs/1506.03447}{{\tt 1506.03447}}].

\bibitem{Palti:2015xra}
E.~Palti, \emph{{On Natural Inflation and Moduli Stabilisation in String
  Theory}}, \href{http://dx.doi.org/10.1007/JHEP10(2015)188}{\emph{JHEP} {\bf
  10} (2015) 188}, [\href{http://arxiv.org/abs/1508.00009}{{\tt 1508.00009}}].

\bibitem{Kooner:2015rza}
K.~Kooner, S.~Parameswaran and I.~Zavala, \emph{{Warping the Weak Gravity
  Conjecture}},
  \href{http://dx.doi.org/10.1016/j.physletb.2016.05.082}{\emph{Phys. Lett.}
  {\bf B759} (2016) 402--409}, [\href{http://arxiv.org/abs/1509.07049}{{\tt
  1509.07049}}].

\bibitem{kappl:2015esy}
R.~Kappl, H.~P. Nilles and M.~W. Winkler, \emph{{Modulated Natural Inflation}},
  \href{http://dx.doi.org/10.1016/j.physletb.2015.12.073}{\emph{Phys. Lett.}
  {\bf B753} (2016) 653--659}, [\href{http://arxiv.org/abs/1511.05560}{{\tt
  1511.05560}}].

\bibitem{choi:2015aem}
K.~Choi and H.~Kim, \emph{{Aligned natural inflation with modulations}},
  \href{http://dx.doi.org/10.1016/j.physletb.2016.05.097}{\emph{Phys. Lett.}
  {\bf B759} (2016) 520--527}, [\href{http://arxiv.org/abs/1511.07201}{{\tt
  1511.07201}}].

\bibitem{Ibanez:2015fcv}
L.~E. Ibanez, M.~Montero, A.~Uranga and I.~Valenzuela, \emph{{Relaxion
  Monodromy and the Weak Gravity Conjecture}},
  \href{http://dx.doi.org/10.1007/JHEP04(2016)020}{\emph{JHEP} {\bf 04} (2016)
  020}, [\href{http://arxiv.org/abs/1512.00025}{{\tt 1512.00025}}].

\bibitem{Hebecker:2015zss}
A.~Hebecker, F.~Rompineve and A.~Westphal, \emph{{Axion Monodromy and the Weak
  Gravity Conjecture}},
  \href{http://dx.doi.org/10.1007/JHEP04(2016)157}{\emph{JHEP} {\bf 04} (2016)
  157}, [\href{http://arxiv.org/abs/1512.03768}{{\tt 1512.03768}}].

\bibitem{conlon:2016aea}
J.~P. Conlon and S.~Krippendorf, \emph{{Axion decay constants away from the
  lamppost}}, \href{http://dx.doi.org/10.1007/JHEP04(2016)085}{\emph{JHEP} {\bf
  04} (2016) 085}, [\href{http://arxiv.org/abs/1601.00647}{{\tt 1601.00647}}].

\bibitem{baume:2016psm}
F.~Baume and E.~Palti, \emph{{Backreacted Axion Field Ranges in String
  Theory}}, \href{http://dx.doi.org/10.1007/JHEP08(2016)043}{\emph{JHEP} {\bf
  08} (2016) 043}, [\href{http://arxiv.org/abs/1602.06517}{{\tt 1602.06517}}].

\bibitem{Heidenreich:2016jrl}
B.~Heidenreich, M.~Reece and T.~Rudelius, \emph{{Axion Experiments to Algebraic
  Geometry: Testing Quantum Gravity via the Weak Gravity Conjecture}},
  \href{http://dx.doi.org/10.1142/S0218271816430057}{\emph{Int. J. Mod. Phys.}
  {\bf D25} (2016) 1643005}, [\href{http://arxiv.org/abs/1605.05311}{{\tt
  1605.05311}}].

\bibitem{cheung:2014vva}
C.~Cheung and G.~N. Remmen, \emph{{Naturalness and the Weak Gravity
  Conjecture}},
  \href{http://dx.doi.org/10.1103/PhysRevLett.113.051601}{\emph{Phys.Rev.Lett.}
  {\bf 113} (2014) 051601}, [\href{http://arxiv.org/abs/1402.2287}{{\tt
  1402.2287}}].

\bibitem{Liam}
B.~Heidenreich, C.~Long, L.~McAllister, M.~Reece, T.~Rudelius and J.~Stout,
  \emph{Work in progress},  2016.

\bibitem{Montero:2016tif}
M.~Montero, G.~Shiu and P.~Soler, \emph{{The Weak Gravity Conjecture in three
  dimensions}}, \href{http://dx.doi.org/10.1007/JHEP10(2016)159}{\emph{JHEP}
  {\bf 10} (2016) 159}, [\href{http://arxiv.org/abs/1606.08438}{{\tt
  1606.08438}}].

\bibitem{Alim2017}
M.~Alim, I.~Garc{\'i}a-Etxebarria, B.~Heidenreich, M.~Reece and T.~Rudelius. To
  appear.

\bibitem{Fischer1975}
A.~E. Fischer and J.~A. Wolf, \emph{{The Structure of Compact Ricci-Flat
  Riemannian Manifolds}}, {\emph{Journal of Differential Geometry} {\bf 10}
  (1975) 277--288}.

\bibitem{Maldacena:2000mw}
J.~M. Maldacena and C.~Nunez, \emph{{Supergravity description of field theories
  on curved manifolds and a no go theorem}},
  \href{http://dx.doi.org/10.1142/S0217751X01003935,
  10.1142/S0217751X01003937}{\emph{Int. J. Mod. Phys.} {\bf A16} (2001)
  822--855}, [\href{http://arxiv.org/abs/hep-th/0007018}{{\tt
  hep-th/0007018}}].

\bibitem{Benjamin:2016fhe}
N.~Benjamin, E.~Dyer, A.~L. Fitzpatrick and S.~Kachru, \emph{{Universal Bounds
  on Charged States in 2d CFT and 3d Gravity}},
  \href{http://dx.doi.org/10.1007/JHEP08(2016)041}{\emph{JHEP} {\bf 08} (2016)
  041}, [\href{http://arxiv.org/abs/1603.09745}{{\tt 1603.09745}}].

\bibitem{Schwimmer:1986mf}
A.~Schwimmer and N.~Seiberg, \emph{{Comments on the N=2, N=3, N=4
  Superconformal Algebras in Two-Dimensions}},
  \href{http://dx.doi.org/10.1016/0370-2693(87)90566-1}{\emph{Phys. Lett.} {\bf
  B184} (1987) 191--196}.

\bibitem{Horowitz:1996nw}
G.~T. Horowitz and J.~Polchinski, \emph{{A Correspondence principle for black
  holes and strings}},
  \href{http://dx.doi.org/10.1103/PhysRevD.55.6189}{\emph{Phys. Rev.} {\bf D55}
  (1997) 6189--6197}, [\href{http://arxiv.org/abs/hep-th/9612146}{{\tt
  hep-th/9612146}}].

\bibitem{Horowitz:1997jc}
G.~T. Horowitz and J.~Polchinski, \emph{{Selfgravitating fundamental strings}},
  \href{http://dx.doi.org/10.1103/PhysRevD.57.2557}{\emph{Phys. Rev.} {\bf D57}
  (1998) 2557--2563}, [\href{http://arxiv.org/abs/hep-th/9707170}{{\tt
  hep-th/9707170}}].

\bibitem{Sen:1994eb}
A.~Sen, \emph{{Black hole solutions in heterotic string theory on a torus}},
  \href{http://dx.doi.org/10.1016/0550-3213(95)00063-X}{\emph{Nucl. Phys.} {\bf
  B440} (1995) 421--440}, [\href{http://arxiv.org/abs/hep-th/9411187}{{\tt
  hep-th/9411187}}].

\bibitem{Aharony:1999ti}
O.~Aharony, S.~S. Gubser, J.~M. Maldacena, H.~Ooguri and Y.~Oz, \emph{{Large N
  field theories, string theory and gravity}},
  \href{http://dx.doi.org/10.1016/S0370-1573(99)00083-6}{\emph{Phys. Rept.}
  {\bf 323} (2000) 183--386}, [\href{http://arxiv.org/abs/hep-th/9905111}{{\tt
  hep-th/9905111}}].

\bibitem{ArkaniHamed:1998rs}
N.~Arkani-Hamed, S.~Dimopoulos and G.~R. Dvali, \emph{{The Hierarchy problem
  and new dimensions at a millimeter}},
  \href{http://dx.doi.org/10.1016/S0370-2693(98)00466-3}{\emph{Phys. Lett.}
  {\bf B429} (1998) 263--272}, [\href{http://arxiv.org/abs/hep-ph/9803315}{{\tt
  hep-ph/9803315}}].

\bibitem{ArkaniHamed:1998nn}
N.~Arkani-Hamed, S.~Dimopoulos and G.~R. Dvali, \emph{{Phenomenology,
  astrophysics and cosmology of theories with submillimeter dimensions and TeV
  scale quantum gravity}},
  \href{http://dx.doi.org/10.1103/PhysRevD.59.086004}{\emph{Phys. Rev.} {\bf
  D59} (1999) 086004}, [\href{http://arxiv.org/abs/hep-ph/9807344}{{\tt
  hep-ph/9807344}}].

\bibitem{Banks:1999gd}
T.~Banks and W.~Fischler, \emph{{A Model for high-energy scattering in quantum
  gravity}},  \href{http://arxiv.org/abs/hep-th/9906038}{{\tt hep-th/9906038}}.

\bibitem{Giddings:2001bu}
S.~B. Giddings and S.~D. Thomas, \emph{{High-energy colliders as black hole
  factories: The End of short distance physics}},
  \href{http://dx.doi.org/10.1103/PhysRevD.65.056010}{\emph{Phys. Rev.} {\bf
  D65} (2002) 056010}, [\href{http://arxiv.org/abs/hep-ph/0106219}{{\tt
  hep-ph/0106219}}].

\bibitem{Banks:2003vp}
T.~Banks, \emph{{A Critique of pure string theory: Heterodox opinions of
  diverse dimensions}},  \href{http://arxiv.org/abs/hep-th/0306074}{{\tt
  hep-th/0306074}}.

\bibitem{ArkaniHamed:2005yv}
N.~Arkani-Hamed, S.~Dimopoulos and S.~Kachru, \emph{{Predictive landscapes and
  new physics at a TeV}},  \href{http://arxiv.org/abs/hep-th/0501082}{{\tt
  hep-th/0501082}}.

\bibitem{Distler:2005hi}
J.~Distler and U.~Varadarajan, \emph{{Random polynomials and the friendly
  landscape}},  \href{http://arxiv.org/abs/hep-th/0507090}{{\tt
  hep-th/0507090}}.

\bibitem{Dimopoulos:2005ac}
S.~Dimopoulos, S.~Kachru, J.~McGreevy and J.~G. Wacker, \emph{{N-flation}},
  \href{http://dx.doi.org/10.1088/1475-7516/2008/08/003}{\emph{JCAP} {\bf 0808}
  (2008) 003}, [\href{http://arxiv.org/abs/hep-th/0507205}{{\tt
  hep-th/0507205}}].

\bibitem{Dvali:2007hz}
G.~Dvali, \emph{{Black Holes and Large N Species Solution to the Hierarchy
  Problem}}, \href{http://dx.doi.org/10.1002/prop.201000009}{\emph{Fortsch.
  Phys.} {\bf 58} (2010) 528--536}, [\href{http://arxiv.org/abs/0706.2050}{{\tt
  0706.2050}}].

\bibitem{Dvali:2007wp}
G.~Dvali and M.~Redi, \emph{{Black Hole Bound on the Number of Species and
  Quantum Gravity at LHC}},
  \href{http://dx.doi.org/10.1103/PhysRevD.77.045027}{\emph{Phys. Rev.} {\bf
  D77} (2008) 045027}, [\href{http://arxiv.org/abs/0710.4344}{{\tt
  0710.4344}}].

\bibitem{Wagner:2012ui}
T.~A. Wagner, S.~Schlamminger, J.~H. Gundlach and E.~G. Adelberger,
  \emph{{Torsion-balance tests of the weak equivalence principle}},
  \href{http://dx.doi.org/10.1088/0264-9381/29/18/184002}{\emph{Class. Quant.
  Grav.} {\bf 29} (2012) 184002}, [\href{http://arxiv.org/abs/1207.2442}{{\tt
  1207.2442}}].

\bibitem{Heeck:2014zfa}
J.~Heeck, \emph{{Unbroken $B - L$ symmetry}},
  \href{http://dx.doi.org/10.1016/j.physletb.2014.10.067}{\emph{Phys. Lett.}
  {\bf B739} (2014) 256--262}, [\href{http://arxiv.org/abs/1408.6845}{{\tt
  1408.6845}}].

\bibitem{Buen-Abad:2015ova}
M.~A. Buen-Abad, G.~Marques-Tavares and M.~Schmaltz, \emph{{Non-Abelian dark
  matter and dark radiation}},
  \href{http://dx.doi.org/10.1103/PhysRevD.92.023531}{\emph{Phys. Rev.} {\bf
  D92} (2015) 023531}, [\href{http://arxiv.org/abs/1505.03542}{{\tt
  1505.03542}}].

\bibitem{Lesgourgues:2015wza}
J.~Lesgourgues, G.~Marques-Tavares and M.~Schmaltz, \emph{{Evidence for dark
  matter interactions in cosmological precision data?}},
  \href{http://dx.doi.org/10.1088/1475-7516/2016/02/037}{\emph{JCAP} {\bf 1602}
  (2016) 037}, [\href{http://arxiv.org/abs/1507.04351}{{\tt 1507.04351}}].

\bibitem{Cyr-Racine:2015ihg}
F.-Y. Cyr-Racine, K.~Sigurdson, J.~Zavala, T.~Bringmann, M.~Vogelsberger and
  C.~Pfrommer, \emph{{ETHOS---an effective theory of structure formation: From
  dark particle physics to the matter distribution of the Universe}},
  \href{http://dx.doi.org/10.1103/PhysRevD.93.123527}{\emph{Phys. Rev.} {\bf
  D93} (2016) 123527}, [\href{http://arxiv.org/abs/1512.05344}{{\tt
  1512.05344}}].

\bibitem{Saraswat:2016eaz}
P.~Saraswat, \emph{{Weak gravity conjecture and effective field theory}},
  \href{http://dx.doi.org/10.1103/PhysRevD.95.025013}{\emph{Phys. Rev.} {\bf
  D95} (2017) 025013}, [\href{http://arxiv.org/abs/1608.06951}{{\tt
  1608.06951}}].

\bibitem{Kaplan:2015fuy}
D.~E. Kaplan and R.~Rattazzi, \emph{{Large field excursions and approximate
  discrete symmetries from a clockwork axion}},
  \href{http://dx.doi.org/10.1103/PhysRevD.93.085007}{\emph{Phys. Rev.} {\bf
  D93} (2016) 085007}, [\href{http://arxiv.org/abs/1511.01827}{{\tt
  1511.01827}}].

\bibitem{voisin2004integral}
C.~Voisin, \emph{On integral hodge classes on uniruled or calabi-yau
  threefolds}, {\emph{Advanced Studies in Pure Mathematics} {\bf 45} (2006)
  43--73}, [\href{http://arxiv.org/abs/math/0412279}{{\tt math/0412279}}].

\bibitem{kats:2006xp}
Y.~Kats, L.~Motl and M.~Padi, \emph{{Higher-order corrections to mass-charge
  relation of extremal black holes}},
  \href{http://dx.doi.org/10.1088/1126-6708/2007/12/068}{\emph{JHEP} {\bf 12}
  (2007) 068}, [\href{http://arxiv.org/abs/hep-th/0606100}{{\tt
  hep-th/0606100}}].

\bibitem{Brown:2015bva}
A.~R. Brown, D.~A. Roberts, L.~Susskind, B.~Swingle and Y.~Zhao,
  \emph{{Holographic Complexity Equals Bulk Action?}},
  \href{http://dx.doi.org/10.1103/PhysRevLett.116.191301}{\emph{Phys. Rev.
  Lett.} {\bf 116} (2016) 191301}, [\href{http://arxiv.org/abs/1509.07876}{{\tt
  1509.07876}}].

\bibitem{Brown:2015lvg}
A.~R. Brown, D.~A. Roberts, L.~Susskind, B.~Swingle and Y.~Zhao,
  \emph{{Complexity, action, and black holes}},
  \href{http://dx.doi.org/10.1103/PhysRevD.93.086006}{\emph{Phys. Rev.} {\bf
  D93} (2016) 086006}, [\href{http://arxiv.org/abs/1512.04993}{{\tt
  1512.04993}}].

\end{thebibliography}\endgroup

\end{document}